\documentclass[apj]{emulateapj}

\usepackage{rotating}
\usepackage{graphicx}
\usepackage{epstopdf}
\usepackage{amsmath}

\newcommand{\enzo}{{\it Enzo}}
\newcommand{\he}{HE87}
\newcommand{\gmf}{{\it GravitatingMassField}}

\DeclareMathOperator{\sinc}{sinc}

\begin{document}

\title{An adaptive particle-mesh gravity solver for ENZO}
\author{Jean-Claude Passy$^1$, Greg L. Bryan$^2$}

\altaffiltext{1}{Argelander Institute f\"{u}r Astronomie, Bonn Universit\"{a}t, Bonn, Germany}
\altaffiltext{2}{Department of Astronomy, Columbia University, New York, NY, USA}

\begin{abstract}

We describe and implement an adaptive particle-mesh algorithm to solve the Poisson equation 
for grid-based hydrodynamics codes with nested grids. The algorithm is 
implemented and extensively tested within the astrophysical code \enzo\
against the multigrid solver available by default.
We find that while both algorithms show similar accuracy for
smooth mass distributions, the adaptive particle-mesh algorithm 
is more accurate for the case of point masses, and is generally
less noisy.  We also demonstrate that the 
two-body problem can be solved accurately in a configuration with nested grids.
In addition, we discuss the effect of subcycling, and demonstrate that evolving all the levels with the 
same timestep yields even greater precision.

\end{abstract}


\section{Introduction}
\label{sec:intro}

One of the challenging aspects of astrophysical simulations is to accurately and efficiently compute
the gravitational potential $\Phi(\mathbf{r})$ for a given density field $\rho(\mathbf{r})$.
For non-trivial cases, it involves solving the well-known Poisson equation:

\begin{equation}
	\Delta \Phi = 4 \pi G \rho,
\label{eq:PoissonEquation}
\end{equation}

\noindent which for simplicity is often solved in Fourier space:

\begin{equation}
	\hat{\phi}(\mathbf{k}) = \hat{G}(\mathbf{k}) \hat{\rho}(\mathbf{k}),
	\label{eq:PoissonFourier}
\end{equation}

\noindent where $\mathbf{k} = (k_1,k_2,k_3)$ is the wavenumber, and $\hat{G}(\mathbf{k})$,
$\hat{\phi}$, and $\hat{\rho}$ are the Fourier transform of the Green's function, the potential,
and the density, respectively.  Within the context of grid-based methods, 
adaptive mesh refinement (AMR) can be crucial because of the large spatial ranges
covered by self-gravitating systems. 
Whilst solving Equations~(\ref{eq:PoissonEquation})~and~(\ref{eq:PoissonFourier})
is relatively trivial in uniform grid situations, 
it becomes much more difficult in multi-level nested simulations.

Many techniques have been proposed to compute the gravitational potential in such configurations
(see \citealt{2005CSci...88.1088B} for a review).  This includes tree-based method \citep[e.g.,][]{1986Natur.324..446B,
1989ApJS...71..871J}, basis function expansions for specific geometries \citep[e.g.,][]{1995CoPhC..89...45M,
1992ApJ...386..375H}.  Here we focus on hierarchical mesh methods \citep[e.g.,][]{1989ApJS...71..407V,
1995MNRAS.274..287S, 1997ApJS..111...73K, 2001MNRAS.325..845K, 2008ApJS..176..293R}.

The \enzo\ code is an open-source\footnote{http://enzo-project.org} multi-dimensional hydrodynamics 
and $N$-body grid-based code with AMR \citep{1995CoPhC..89..149B,2004astro.ph..3044O,2014ApJS..211...19B}.
In the \enzo\ code, gravity is solved using a particular method, which uses multigrid
\citep{Brandt:1977ve} on refined patches.  This method is fast, but has a number of
shortcomings in certain situations, as we will demonstrate in this paper.
An an alternative to that method, we present another method based 
on the adaptive particle-mesh (APM) algorithm derived from 
\cite{Hockney:1989um} (hereafter \he), and
\cite{1991ApJ...368L..23C},
which allows a more accurate calculation of the potential
when point potentials are employed. We implement this APM technique within the \enzo\ code
and compare its accuracy with the multigrid solver available by default.
In Section~\ref{sec:method} we describe the different algorithms for the multigrid and the APM gravity solvers. 
Several test cases are performed and analyzed in Section~\ref{sec:tests} in order to compare both techniques.
We summarize our results and conclude in Section~\ref{sec:conclusion}.

\section{Numerical technique}
\label{sec:method}

To calculate the acceleration of the gas and the particles due to self-gravity, 
one must first compute the total gravitating field of the system.
The particles are deposited into the $n$ nearest cells using either a cloud-in-cell (CIC, $n=2^3$) for the multigrid solver,
or a triangular-shape cloud (TSC, $n=3^3$) interpolation technique for the APM solver.
For the CIC deposition we use a cloud size equal to the target grid size, therefore switch the value 
of the parameter {\it ParticleSubgridDepositMode} to 0.
Note that both CIC and TSC algorithms are equivalent for a particle located at a corner of a cell.

We have slightly modified the deposition algorithm for the particles. 
Normally, particles that belong to a given grid or to any of its children are deposited onto that grid directly.
More specifically, the particles density is copied into its \gmf\
which contains the density contribution from the particles and the baryons. 
The \gmf\ is larger that the grid, therefore can have boundary points set by the parent grid.
Thus we eventually need to copy the \gmf\ of the parent grid onto the grid boundary 
(for more details, see \citealt{2014ApJS..211...19B}).
We have found that this scheme could lead to the contribution from a particle located near the grid edges, 
being accounted twice -- once from the grid itself and once from the parent grid.
Therefore we instead deposit the particles from the parent grid and the parent siblings also, 
and then copy the baryonic contribution from the parent grid only onto the grid boundary.

Before mass deposition, the particles are advanced by half a timestep in order to obtain a time-centered density field.
The density field generated by the particles is then added to the gas density of the cells which are also time-centered.
At this point, the total gravitating mass field has been calculated and will be used to compute the gravitational potential.
We describe this step for both solvers in the next two sections.

\subsection{The multigrid solver}
\label{sub:multigrid}

The default gravity solver implemented in \enzo\ is based on a combination of Fourier and 
multigrid algorithms. 
We recall here the basics of this method, but for more details, see \cite{2014ApJS..211...19B}.
First the gravitating potential is computed on the root grid in Fourier space
using a fast Fourier transform (FFT, \he).
For periodic boundary conditions, the Green's function is generated directly in Fourier-space.
For isolated boundary conditions, it is calculated first in real space to obtain the correct zero-padding, and transformed in the Fourier domain \citep{James:1977dn}.
In both cases, the resulting potential is then transformed to the real domain and differentiated in order to obtain 
the face-centered (cell-centered) acceleration field depending on whether 
the hydro solver requires face-center or cell-centered accelerations.
On the subgrids, boundary conditions for the potential are obtained by interpolation from the parent grid
and then Equation~(\ref{eq:PoissonEquation}) is solved using a Gauss-Seidel multigrid relaxation method with 
Dirichlet boundary conditions (\he).
As we will highlight in Section~\ref{sec:tests}, such algorithm may introduce errors
on the refined grids due to inaccuracies in the coarse-grid solution.
We should also emphasize that the particular multigrid solver described here is not strictly speaking a pure multigrid
solver in the traditional sense.
Indeed, classic multigrid solvers calculate the solution on all levels at the same time.

\subsection{The APM solver} 
\label{sub:apm}

The APM solver is based on the particle-particle adaptive particle-mesh method 
\citep[\he,][]{1991ApJ...368L..23C}. The general idea of the algorithm is to split
the gravitational force between a long-range component, 
and one or more short-range components which are nonzero only for a narrow range of wavenumbers.
The algorithm as described in this section is suitable for three-dimensional problems only, 
but could be extended in the future to treat one- and two-dimensional systems.

The calculation on the root grid is nearly identical to the one performed with the multigrid solver (although
the Greens function is slightly modified). 
On a refined grid one first calculates the short-range component of the force.
Therefore, the Green's function needs to be modified in order to account for the contribution from the smallest scales only.
We thus need a smoothing function, which here is the sphere with uniformly decreasing density $S(r)$: 
\begin{equation}	
	S(r) = \left\{
	\begin{array}{ll}
	\frac{48}{\pi a^4} \left( \frac{a}{2} - r \right) \ \ \ \ \ {\rm if} \ \ \ \ \ r \leq a/2 \\ \\
	0 \hspace{2.3cm} {\rm otherwise,}
	\end{array}
	\right.
	\label{eq:S2}
\end{equation}
where $r$ is the radius and $a$ is a positive parameter. The corresponding Fourier transform is 
\begin{equation}	
	\hat{S}_l (k) = \frac{12}{\eta^4} \left(2 - 2\cos{\eta} - \eta \sin{\eta}\right),
	\label{eq:S2k}
\end{equation}
where $k= |\mathbf{k}|$ and $\eta = k a/2$. 
For each refinement level we take $a = 3.4\delta_l$, where $\delta_l$ is the linear size of a grid cell on the current level $l$.
The effects of the smoothing function and the smoothing parameter have been studied extensively in \he,
who concluded that this profile gives a better accuracy in three-dimensional schemes and
this chosen value for $a$ minimizes the numerical noise.

Minimizing the error in the mesh force leads to the optimal Green's function (\he, Equation~8-22):
\begin{equation}
	\hat{G}(\mathbf{k}) = \frac{\mathbf{\hat{D}} (\mathbf{k})  \cdot \sum_\mathbf{n} \hat{U}^2(\mathbf{k_n}) \mathbf{\hat{R}}(\mathbf{k_n})}
	{|\mathbf{\hat{D}} (\mathbf{k})|^2 \left[  \sum_\mathbf{n} \hat{U}^2(\mathbf{k_n}) \right]^2 },
	\label{eq:influence}
\end{equation}
where 
\begin{equation}
	\mathbf{k_n} = \mathbf{k} +\mathbf{n} 2\pi/\delta_l, \ \mathbf{n} \in [-\infty,+\infty]^3
	\label{eq:brillouin}
\end{equation} 
%
are the Brillouin zones,
$\mathbf{\hat{D}}$ is the gradient operator, $U$ is the assignment function, and $\mathbf{R}$ is the reference force.
For the gradient operator we use either a direct Fourier-space solver ($\hat{D}(\mathbf{k}) = -i \mathbf{k}$) or a finite-difference representation (see \he\ for an explcit expression).
For TSC deposition, the assignment function $U$ is (\he, Equations 8-42,~8-45):
\begin{equation}
	\hat{U}(\mathbf{k})  = \prod_{i=1}^3 \sinc^6\left({\frac{k_i\delta_l}{2}}  \right)
	\label{eq:assignment}
\end{equation}
and verifies
\begin{equation}
	\sum_\mathbf{n} \hat{U}^2(\mathbf{k_n})  = \prod_{i=1}^3 \left( 1-\sin^2{\frac{k_i\delta_l}{2}} + \frac{2}{15} \sin^4{\frac{k_i\delta_l}{2}}  \right).
	\label{eq:assignment2}
\end{equation}
Finally, the reference force is linked to the smoothing function by:
\begin{equation}
	\hat{\mathbf{R}}(\mathbf{k}) = -i \frac{\hat{S}_l^2 - \hat{S}_{l-1}^2 }{k^2} \mathbf{k}.
	\label{eq:reference}
\end{equation}
$\hat{S_l}$ signifies the smoothing function with cell size on level $l\mathbf{\geq1}$.  For the coarsest level ($l=0$), 
this becomes $-i \hat{S_{0}} \mathbf{k} / k^2$.
Note that, since $|\hat{\mathbf{R}}(\mathbf{k})| \propto k^{-9}$ and $|\hat{U}(\mathbf{k})| \propto k^{-3}$,
we only consider the Brillouin zones with $\mathbf{n} \in [-1,1]^3$ in Equation~(\ref{eq:influence}).

The resulting potential is computed using Equations~(\ref{eq:PoissonFourier})~and~(\ref{eq:influence}), 
and then differenced to obtain the acceleration in Fourier space. An inverse FFT is used to obtain the 
mesh force in real space.  Finally, the total acceleration field is computed by interpolating the 
acceleration field from the parent grid onto the refined grid, and adding it to the fine mesh force.
The FFTs used for the refined meshes require zero-padding, but because the short-range force is
compact, we have found that we only require an extra $0.65 \mathcal{R} a/\delta$ cells, where $\mathcal{R}$ is the
usual refinement factor between levels.
In this paper we always use  $\mathcal{R} = 2$.

\begin{figure}
	\begin{center}
		\includegraphics[scale=0.46, trim={0.7cm 0.5cm 0 0}]{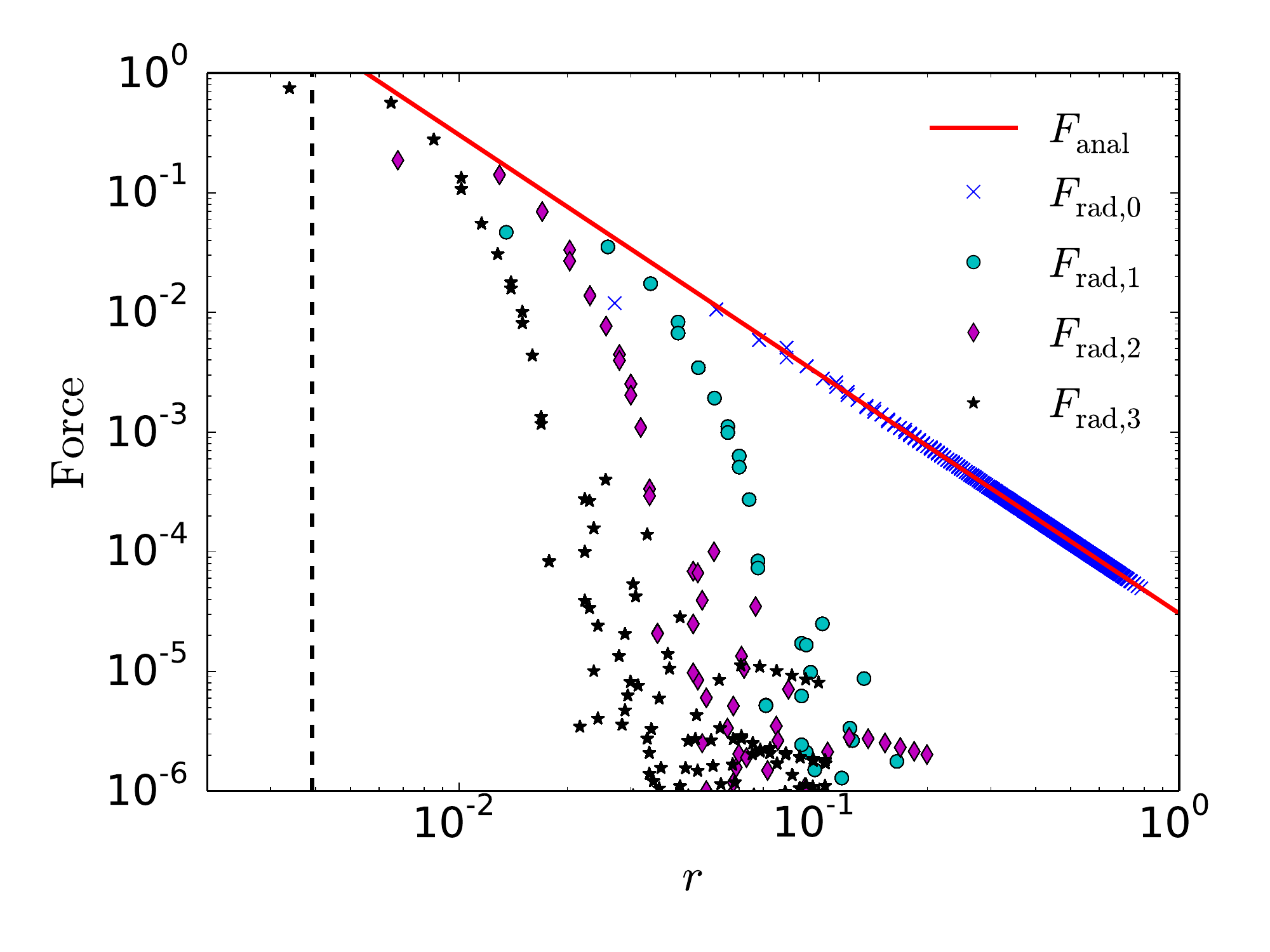}
	\caption{Forces created on the grids by the central particle as a function of the distance between
	a given cell and the particle position.
	Shown are the different radial components computed with the APM solver for each level  
	(colored symbols), and the total analytical force (red solid line).
	The vertical dashed line represents the size of the smallest grid cell $\delta_3 = 3.906\times10^{-3}$.
	\label{fig:apm}
	}
	\end{center}
\end{figure}

For illustrative purposes a decomposition of the gravitational force is shown 
in Figure~\ref{fig:apm} for a particle located at the center of the computational domain. 
The resolution of the root grid is $32^3$ zones, and we add three levels of refinement
with respective coordinates 
[0.1875\,;\,0.8125]$^3$, [0.3125\,;\,0.6875]$^3$, and [0.40625\,;\,0.59375]$^3$.
The components of the force on each level are shown in Figure~\ref{fig:apm}, and compared to the 
analytical force $F_{\rm anal} \propto r^{-2}$, where $r$ is the distance from the central particle. 
We can clearly see how each subgrid contributes 
on the smallest scales only, while the larger scale components are computed on the levels above.
Note that the computed acceleration field is always cell-centered. 
It is therefore interpolated linearly when the Euler equations are solved in case a staggered
mesh hydro scheme is used.

Our implementation within the \enzo\ AMR code includes parallelization through the MPI library.
This is relatively straightforward, as most of the density construction was already parallelized, with
the largest change being communication of the coarse accelerations from the parent grid to the refine
grid.  In addition to parallelization, we have implemented a method to cache frequently 
used Green's functions, so that they do not need to be recomputed for each new grid.

\section{Tests}
\label{sec:tests}

In this section we perform a series of tests in order to compare both gravity solvers in terms of accuracy.
All the tests are performed in parallel on four cores using 
our modified version of \enzo\ v2.3
on a MacBook Pro OS X 10.8.5 with GNU compilers (gcc 4.2.1 and gfortran 4.9.0) and OpenMPI 1.6.5.
We use the PPM scheme \citep{1984JCoPh..54..174C} to solve the hydrodynamics equations.
The coarse grid covers the entire three-dimensional computational domain 
which has coordinates [0\,;\,1]$^3$ in code units. Part of the domain is refined using nested grids.
Each subgrid is refined by a factor 2 in comparison with the resolution of its parent grid.
Grids can be therefore connected and organized using a tree where level 0 corresponds to the root grid,
and all grids with the same resolution are located on the same level.
For the tests that only involve particles, it might very well be that a grid does not contain 
any particle, thus preventing us from using the usual Courant conditions to determine
timestepping \cite[for more details, see][]{2014ApJS..211...19B}. 
Therefore we include an additional condition that constrains
the timestep on level $l+1$ to be at most smaller than half of the timestep on level $l$:
\begin{equation}
	dt_{l+1} \leq dt_l/2.
	\label{eq:timestep}
\end{equation}
Note that this condition is automatically fulfilled if gas or particles are present 
in the grid. For all tests except for the sine wave problem (Section~\ref{subsec:sinewave})
we use isolated boundary conditions on the root grid.

In what follows we note $m$ is the mass of a given particle and $\rho = m/ \Delta V$ is its density, 
where $\Delta V$ is the volume of a zone of the grid in which the particle is located.

\subsection{The self force test}
\label{subsec:selforce}

We first verify that there is no self force acting on a particle. This is done by modeling the evolution of an isolated
particle that is given an initial velocity and goes through the different levels of refinement. If the particle does not
experience any self force, it's velocity should not change.

The root grid resolution is $16^3$ zones and there are three static levels of refinement with coordinates 
[0.1875\,;\,0.8125]$^3$, [0.3125\,;\,0.6875]$^3$, and [0.40625\,;\,0.59375]$^3$.
The particle of mass $m = 1$ is initially located on level 0 at coordinates $\mathbf{r}_0 = (0.15, 0.15, 0.15)$, 
and is given a velocity $\mathbf{v}_0 = (1.0, 1.0, 1.0)$. 
The system is evolved until $t = 0.4$, and we investigate two cases.

In the first case, we don't allow any subcycling and all levels are advanced at a constant timestep $dt = 3 \times 10^{-3}$.  
With the APM solver, the particle acceleration is found to be of the order $10^{-14}$ on 
the deepest level, therefore completely negligible. 

\begin{figure}
	\begin{center}
		\includegraphics[scale=0.45, trim={0.7cm 0.5cm 0 0}]{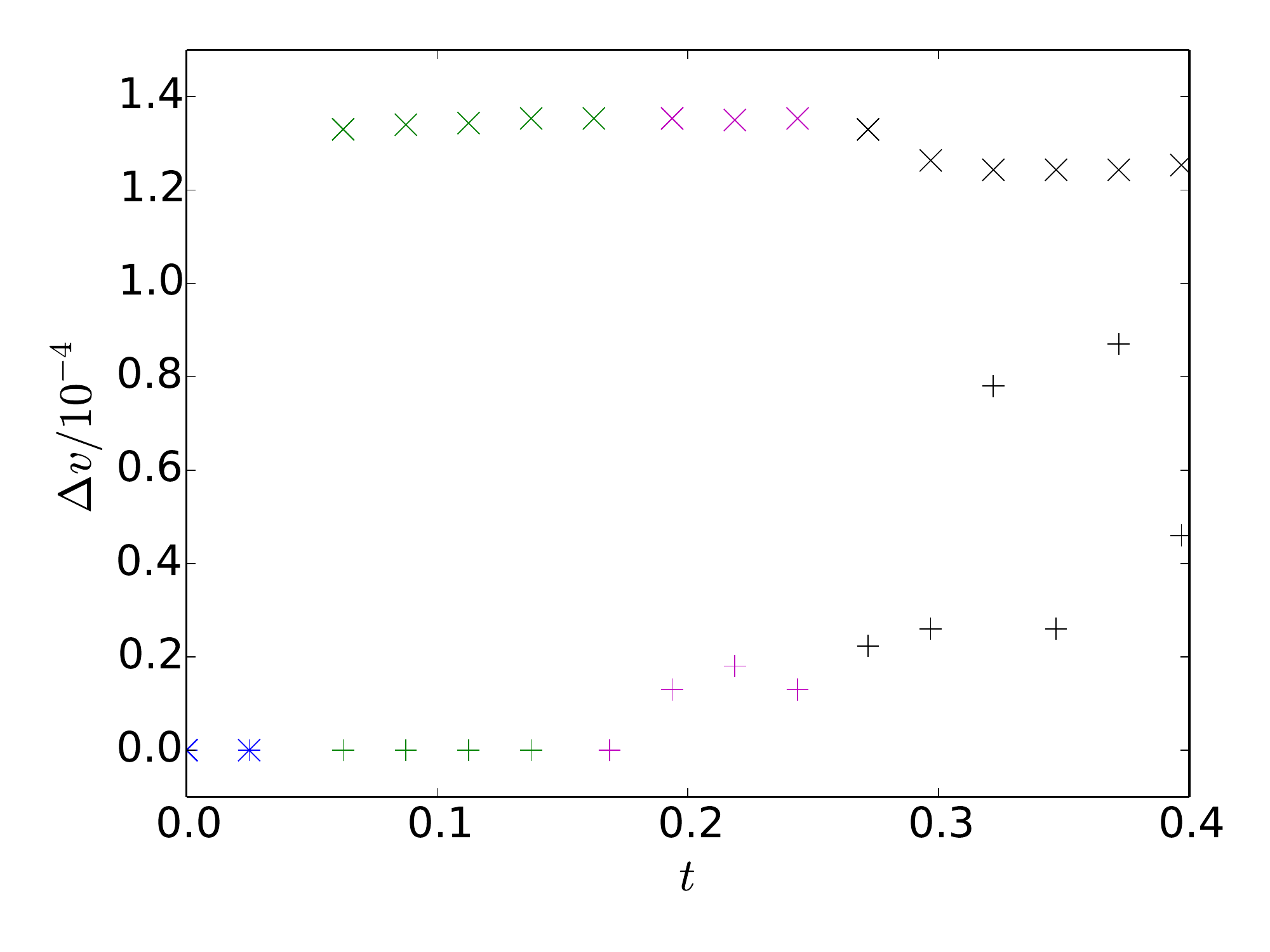}
	\caption{Offset in the particle velocity as a function of time for the multigrid (cross) and the APM (plus) solvers.
	The different colors represent the position of the particle in the hierarchy: level 0 (blue), 1 (green), 2 (purple), and 3 (black). 
	Data points are 
	equally spaced because we plot velocities only at the end of a root grid cycle.  Note that this figure only plots the case when
	subcycling in the time step is used -- if subcycling is not employed, the velocities errors are negligible ($\approx 10^{-14}$).
	\label{fig:selfforce}
	}
	\end{center}
\end{figure}

In the second case, the system is evolved at a fixed timestep $dt_0 = 2.5\,10^{-2}$ on level 0, 
and with subcycles on the refined levels. Using the added criterion for timestepping (Equation~\ref{eq:timestep}),
the value of $dt_0$ leads to the deepest level (level 3) being evolved with a timestep similar to the constant timestep
chosen in the first case.
We show in Figure~\ref{fig:selfforce} the error on the particle velocity 
$\Delta v = |\mathbf{v}(t) - \mathbf{v}_0|/ |\mathbf{v}_0|$ for both solvers.
The error for both methods is still small, of the order $10^{-4}$, 
and the APM solver is slightly more accurate than the multigrid solver.
However, allowing subcycles introduced an error due to the time-integration, 
error that is symmetric in all directions.
We will witness this effect
in Section~\ref{subsec:testorbit} as well.
We thus conclude that the APM solver does not directly introduce any self-force but that evolving the different levels
with different timestepping decreases the time-integration accuracy. 
This is essentially because with subcycling, the force contribution from the different levels are computed at different times.
Figure~\ref{fig:selfforce} also demonstrates that the multigrid solver does introduce self-forces, 
in particular for particles located near the boundary of a refined region.

\subsection{The point source gravity test}
\label{subsec:gravitytest}

In this test we compute the acceleration of nearly massless particles in a gravitational field created by a heavy central particle.
The dimensions of the root grid are $32^3$, and we add one static refined grid.
We set up at the center a particle of density $\rho = 1.0$, and $n =5\,000$ test particles
of mass $m = 10^{-10}$ distributed randomly in $\log{r}$, where $r$ is the distance from the central particle.
We study two cases: one in which the refined grid covers the region [0.4375\,;\,0.5625]$^3$ such that the particle is at the center of 
the subgrid, and one in which the refined region is [0.5\,;\,0.5625]$^3$ such that the particle is deposited on the corner of the subgrid. In both cases we evolve the system for $\Delta t = 10^{-10}$.

\begin{figure*}[!h]
	\begin{center}
		\includegraphics[scale=0.43]{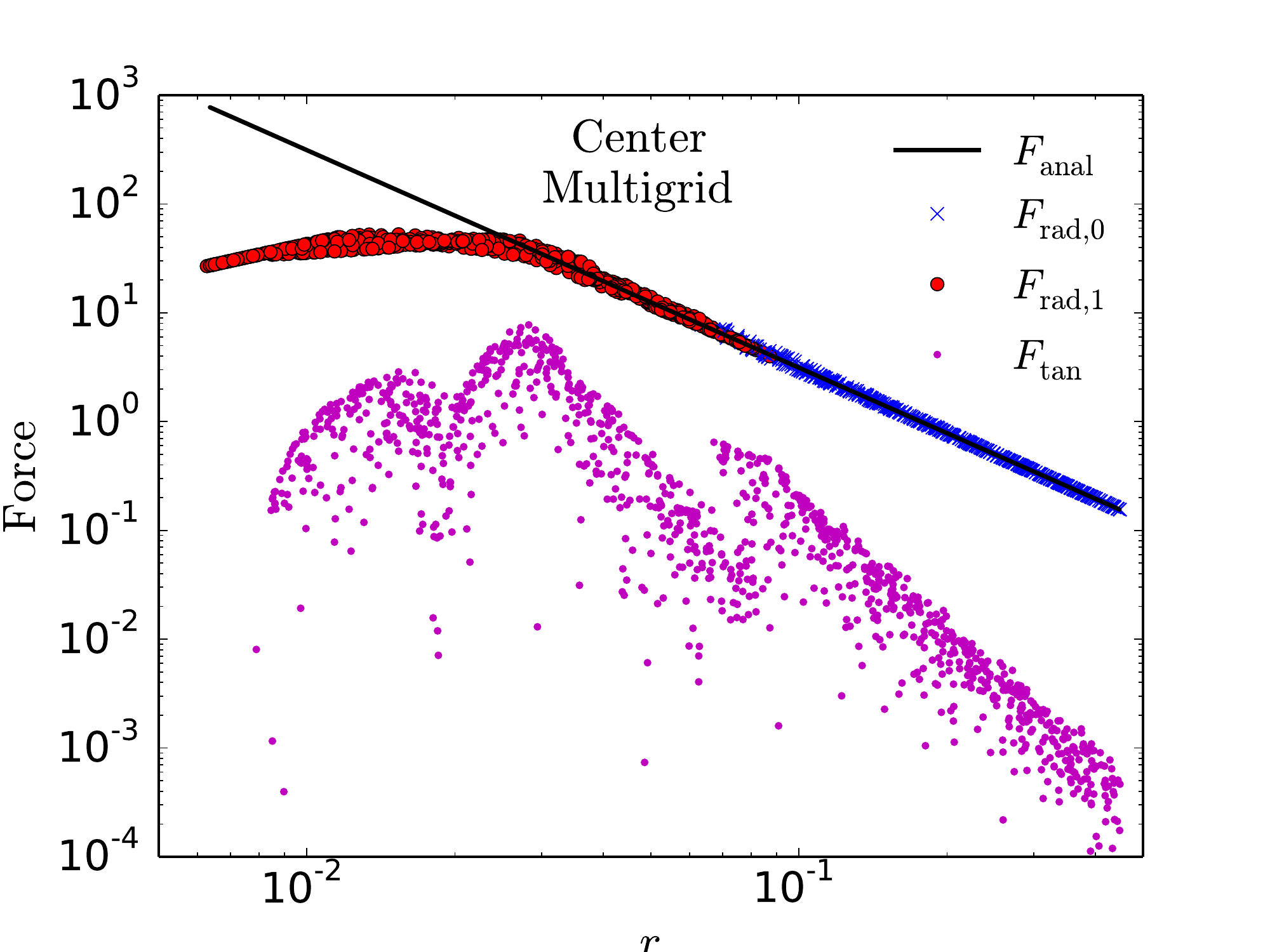}
		\includegraphics[scale=0.43]{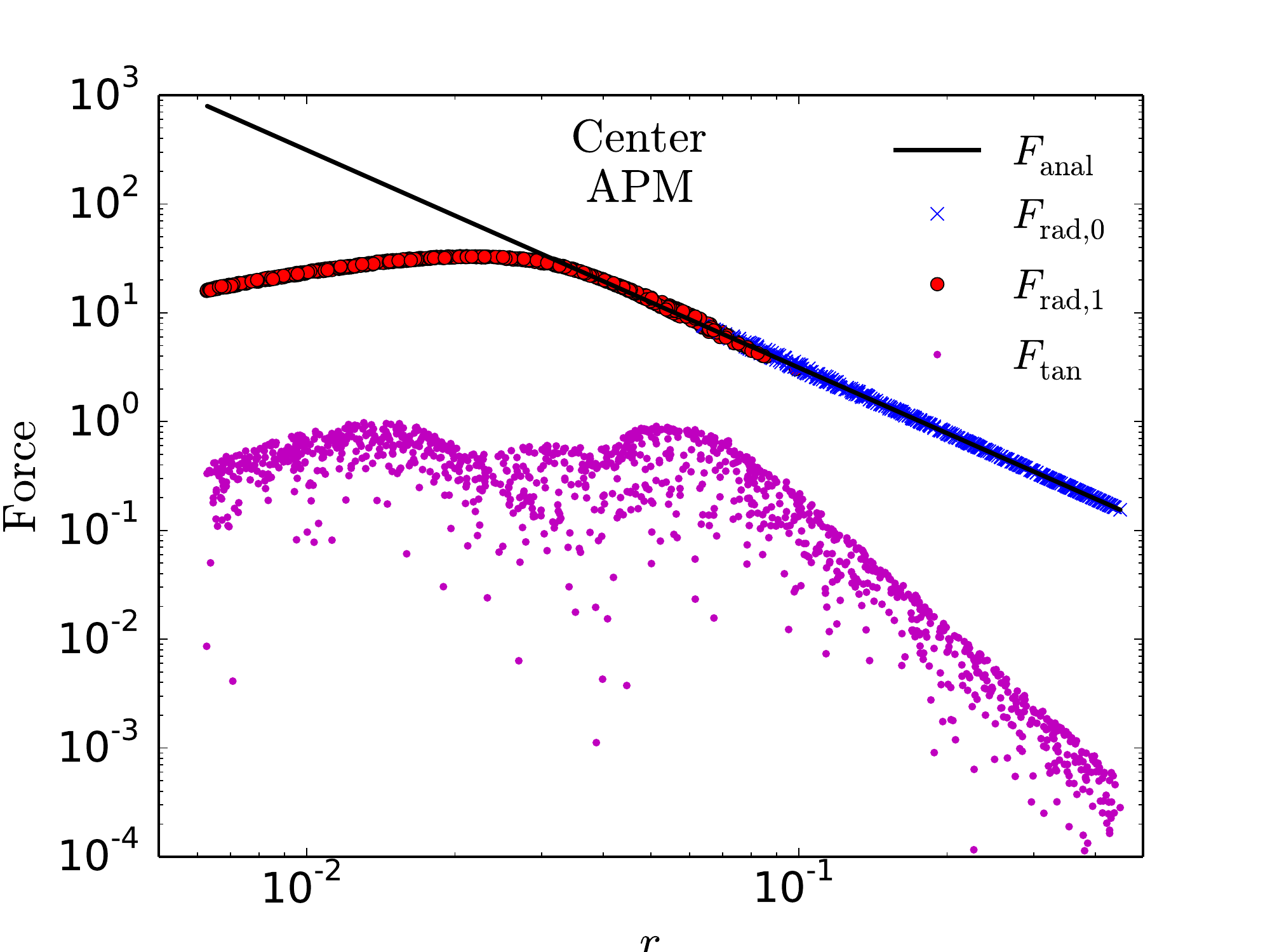}
		\includegraphics[scale=0.43]{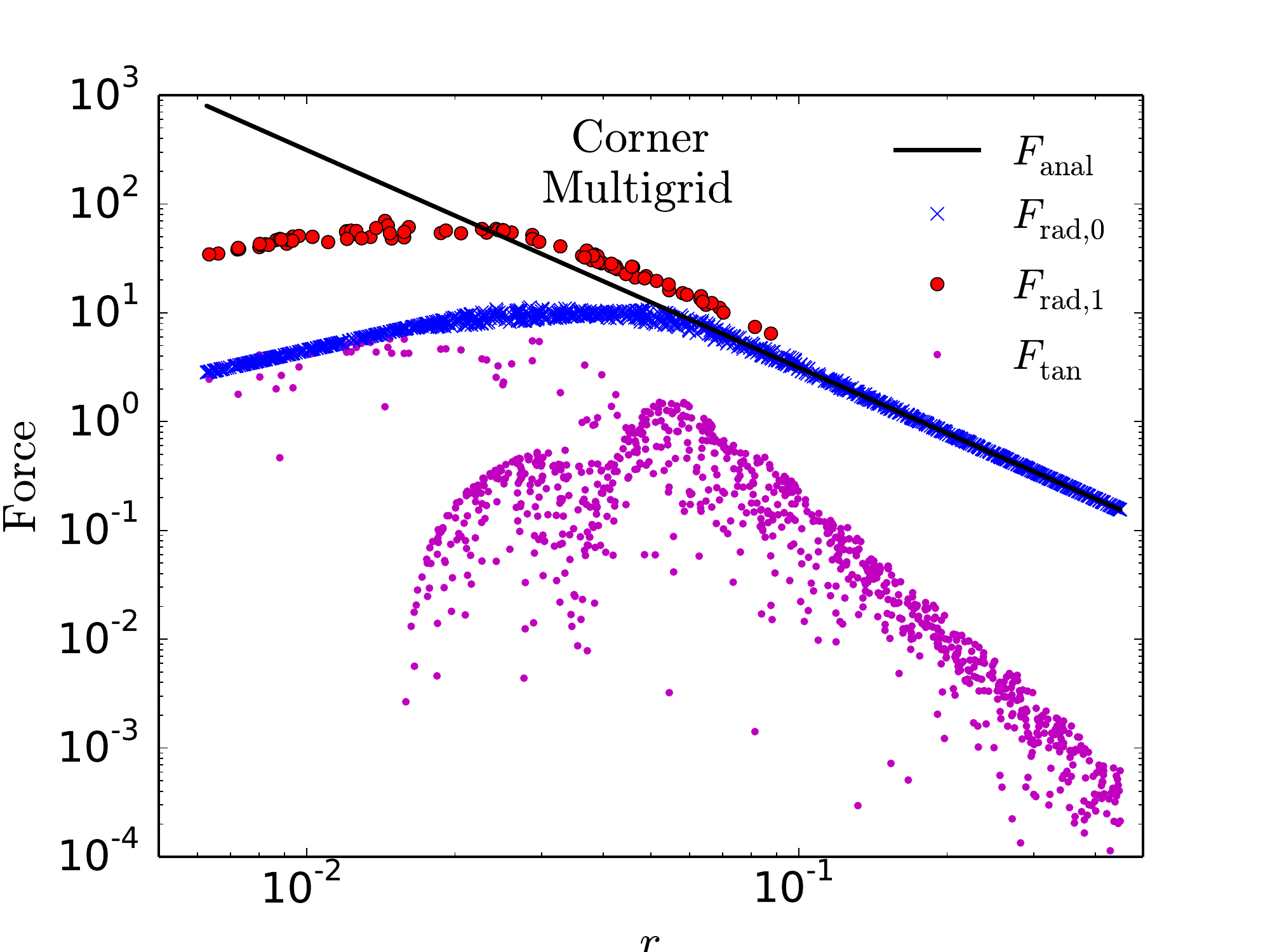}
		\includegraphics[scale=0.43]{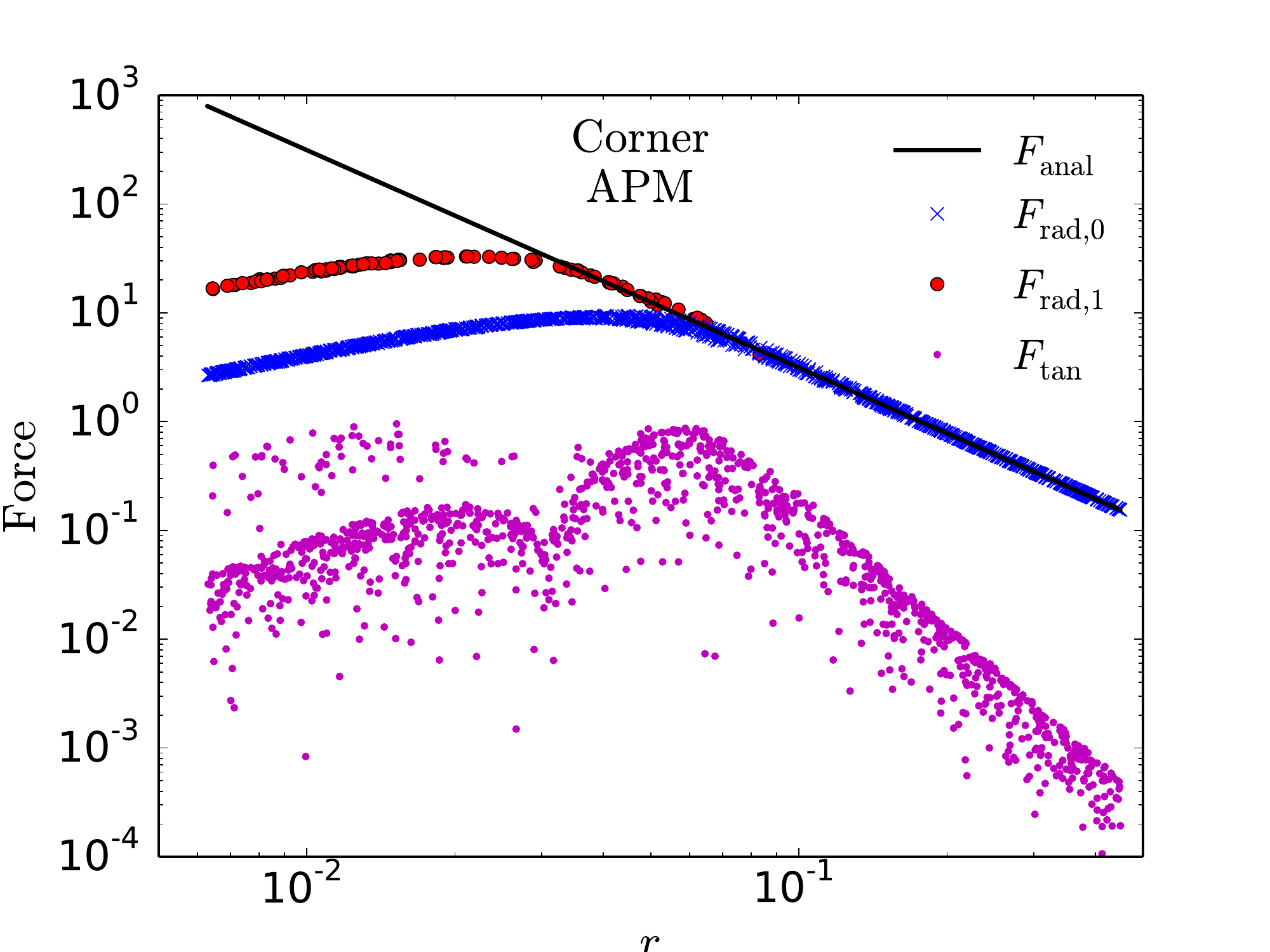}
	\caption{Forces on the test particles as a function of the distance to the 
	central particle obtained 
	with the multigrid (left) and the APM (right) solvers. 
	Plotted are the analytical force (solid black), the computed radial forces for the root grid (blue crosses) 
	and the refined grid (red filled circles), and the computed tangential force (purple dots), for the case where
	the massive particle is located at the center (top) and at a corner (bottom) of the refined grid.
	\label{fig:gravitytest}}
	\end{center}
\end{figure*}

The different force components are plotted in Figure~\ref{fig:gravitytest}.
The tangential component of the force $F_{\rm tan}$ should be zero while the radial component 
$F_{\rm rad}$ should follow the analytical result $F_{\rm anal} \propto r^{-2}$, 
but is softened for radii about two cell lengths. 
The largest inaccuracy in the force calculation 
is reached at the boundary of the nested grid ($r \approx 2$).
Both solvers give an overall good result, 
as the relative mean errors of the radial and tangential accelerations are only of the order of a few percent. 
However, the transition is much smoother with the APM
solver than with the multigrid solver. With the APM solver the overall noise in the radial force 
is reduced in comparison with the multigrid solver (Figure~\ref{fig:gravitytest}). 

Note that in the case where the massive particles is located at the edge of the refined grid, 
one can see how the smoothing of the potential starts at larger distances 
on the root grid than on the refined grid.
Moreover, the acceleration of the central particle in the first case is $\approx 10^{-8}$ and $\approx 10^{-11}$
for the multigrid and the APM solvers, respectively. In the second case, the respective accelerations are 
$\approx 10^{-3}$ and $\approx 10^{-10}$.
We therefore confirm that
there is no self-force introduced by the APM solver, as demonstrated in Section~\ref{subsec:selforce}.

\subsection{The sphere gravity test}
\label{subsec:gravitysphere}

We investigate here the error in a spherical distributions for which we can
analytically calculate the potential. These are the sphere of constant density ($i=1$), the isothermal sphere ($i=2$)
and the Plummer sphere ($i=3$). We note $M$ and $R$ are the total mass and radius of the sphere, respectively.
The density distributions for the three cases are given by:
\begin{equation}
	\rho_i(r) = \left\{
	\begin{array}{ll}
		\rho_0 \hspace{2.5cm} r \leq R, \ i = 1 \\ \\
		\rho_0 \left(\frac{R}{r}\right)^2 \hspace{1.6cm} r \leq R, \ i = 2\\ \\
		\frac{\rho_0}{\left[1 + (r/R)^2\right]^{5/2}} \hspace{1cm} r \leq R, \ i = 3\\ \\ \\ 
		\rho_{{\rm out}} \hspace{2.2cm} r > R 
	\end{array}
	\right.
\end{equation}
where $\rho_0 = 1$ and $\rho_{{\rm out}} = 10^{-10}$.
The corresponding potentials are obtained using Gauss's law:
\begin{equation}
	\Phi_i(r) = \left\{
	\begin{array}{ll}
		-\frac{M}{2R} \left(3-r^2/R^2\right) \hspace{1cm} r \leq R, \ i = 1 \\ \\
		-\frac{M}{R} \left[1-\log{(r/R)}\right] \hspace{0.70cm} r \leq R, \ i = 2\\ \\
		-\frac{M}{R} \left(\frac{2\sqrt{2}}{\sqrt{1+r^2/R^2}}-1\right) \hspace{0.25cm} r \leq R, \ i = 3\\ \\ \\ 
		-\frac{M}{r} \hspace{3cm} r \geq R 
	\end{array}
	\right.
\end{equation}

\begin{figure*}[!h]
	\begin{center}
	\includegraphics[scale=0.495]{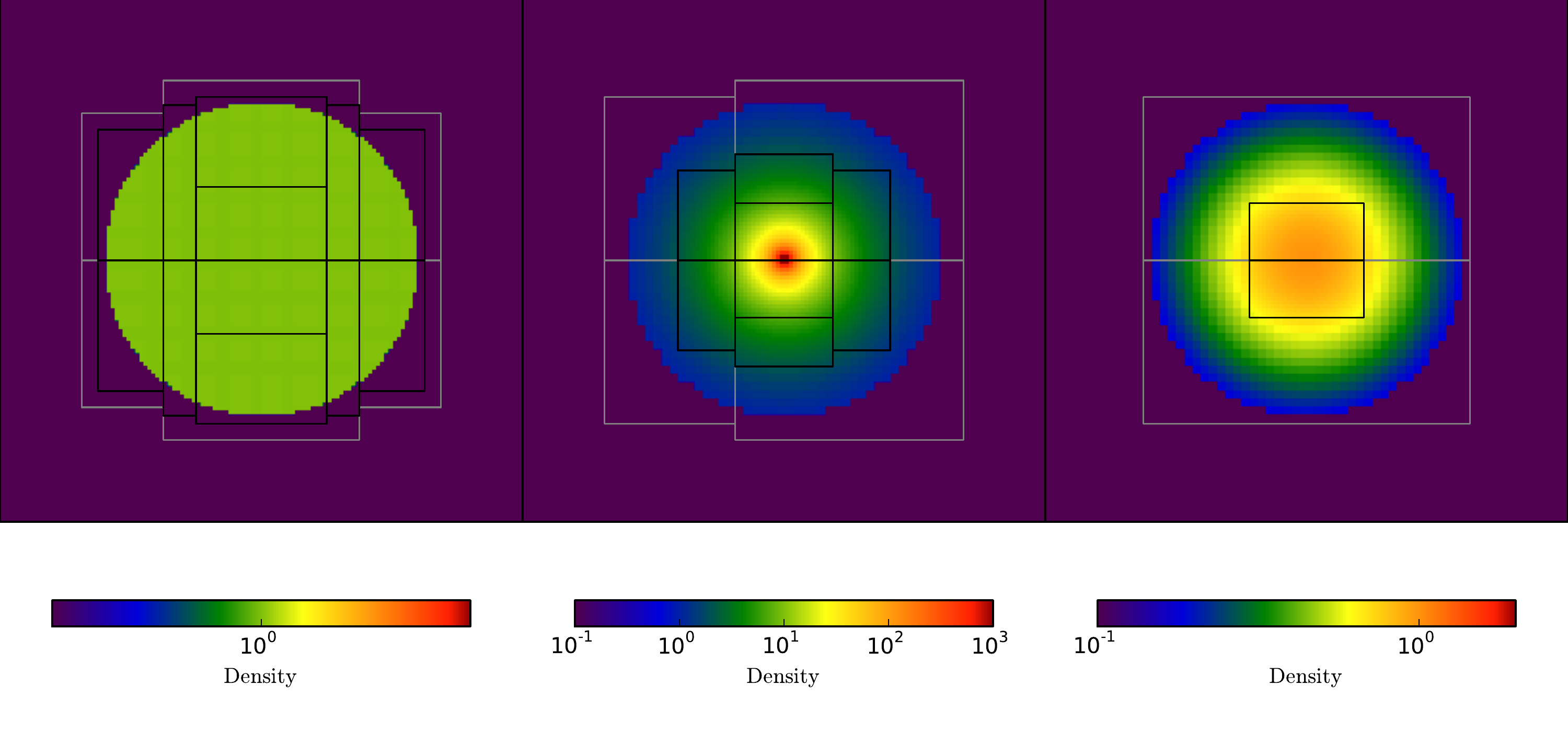}
	\caption{Density slice in the plane $z = 0.5$ for the three cases: 
	the sphere of constant density ($i=0$, left), 
	the isothermal sphere ($i=1$, middle) and the Plummer sphere ($i=2$, right).
	The edges of the subgrids are also shown in gray (level 1) and black (level 2).
	\label{fig:grids_sphere}}
	\end{center}
\end{figure*}

In each case the dimensions of the root grid are $32^3$ and $R = 0.3$.
We allow up to two levels of refinement and 
the domain is refined using the parameter {\it MinimumMassForRefinement} 
which we set to $10^{-6}, 10^{-5}, 3\times10^{-6}$ for $i=1,2,3$.
Refinement occurs if the mass in a cell exceeds this parameter.
The resulting hierarchies are plotted in Figure~\ref{fig:grids_sphere}.

\begin{figure*}[!h]
	\begin{center}
		\includegraphics[scale=0.44]{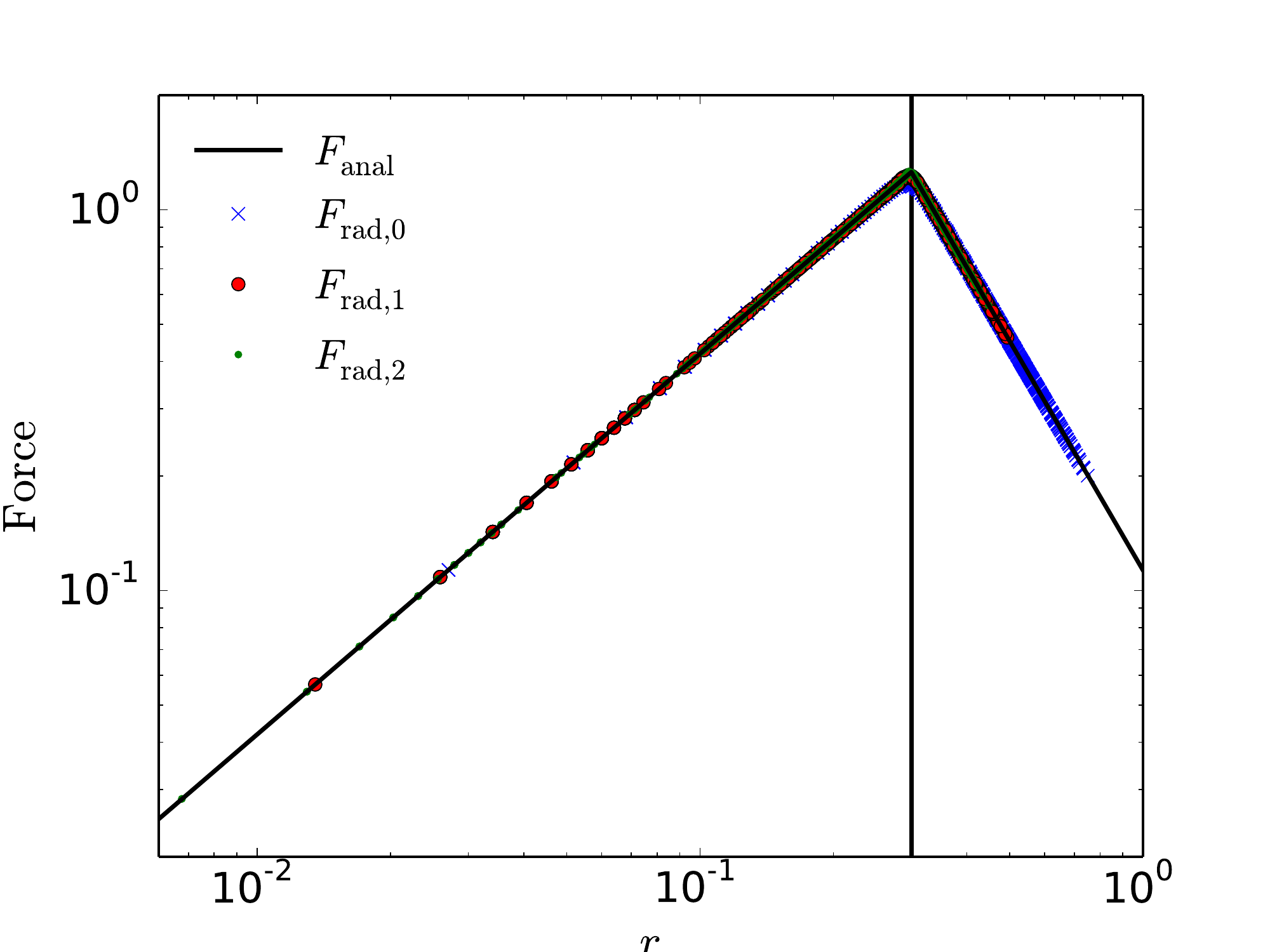}
		\includegraphics[scale=0.44]{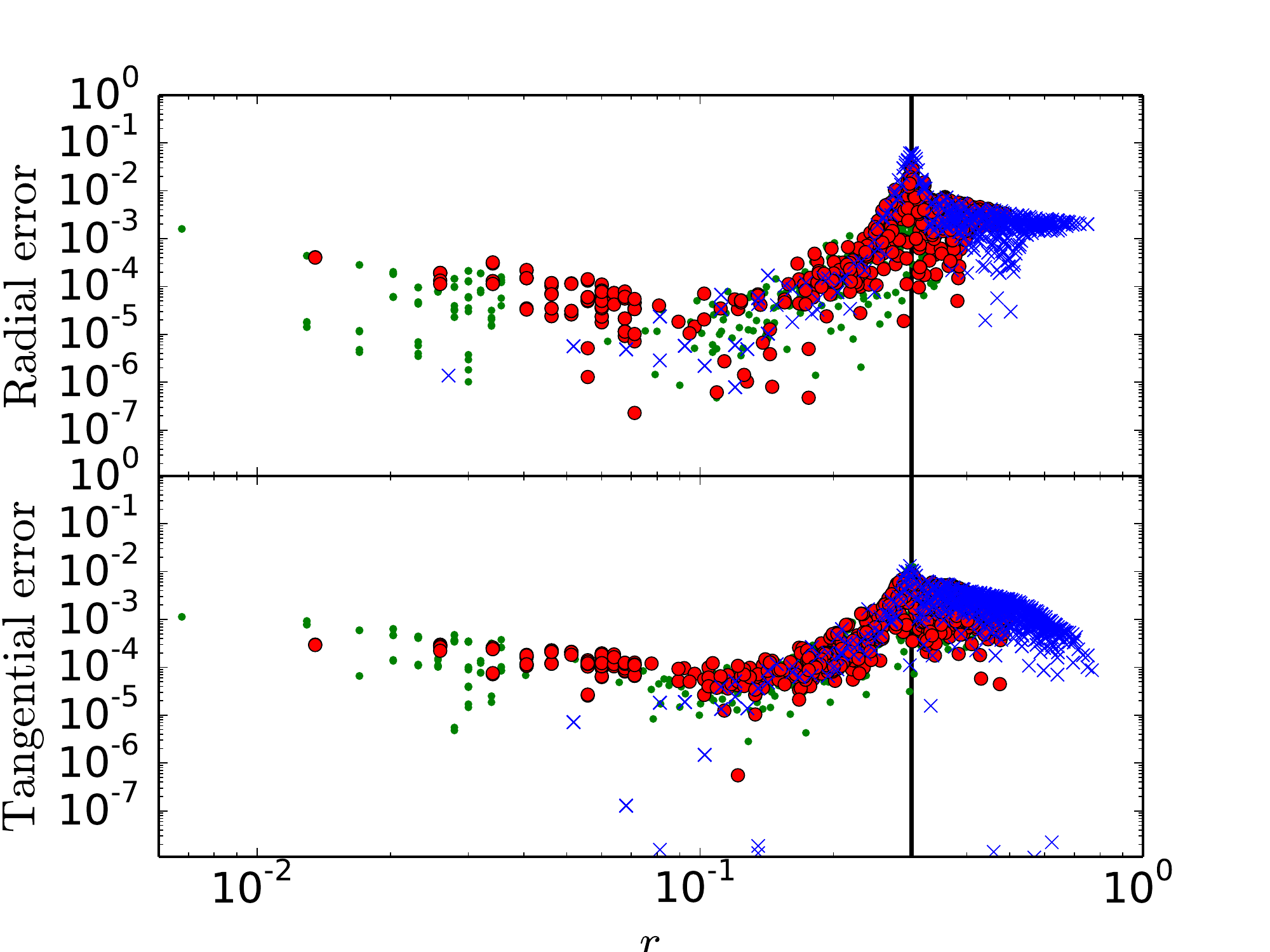}
		\includegraphics[scale=0.44]{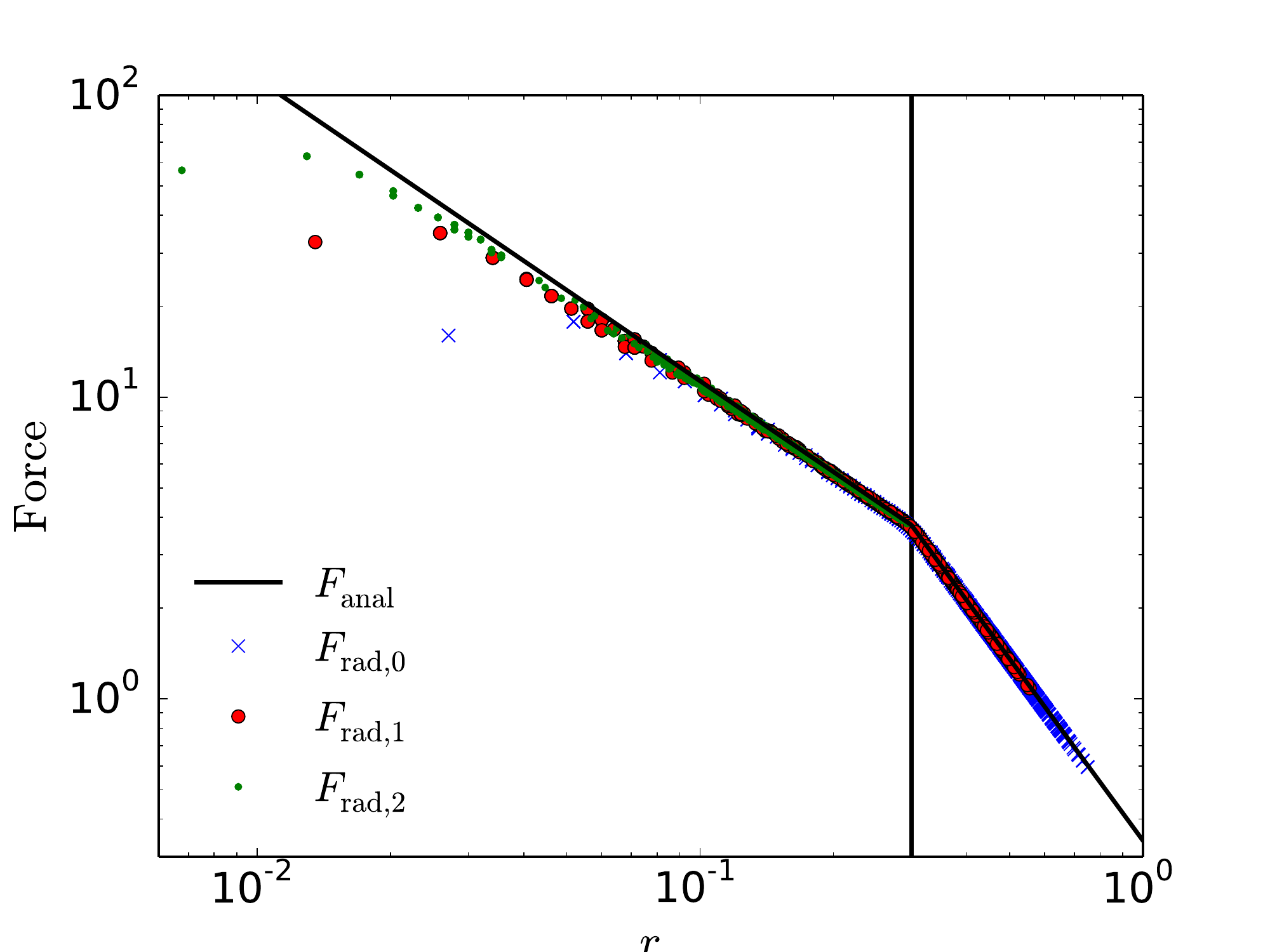}
		\includegraphics[scale=0.44]{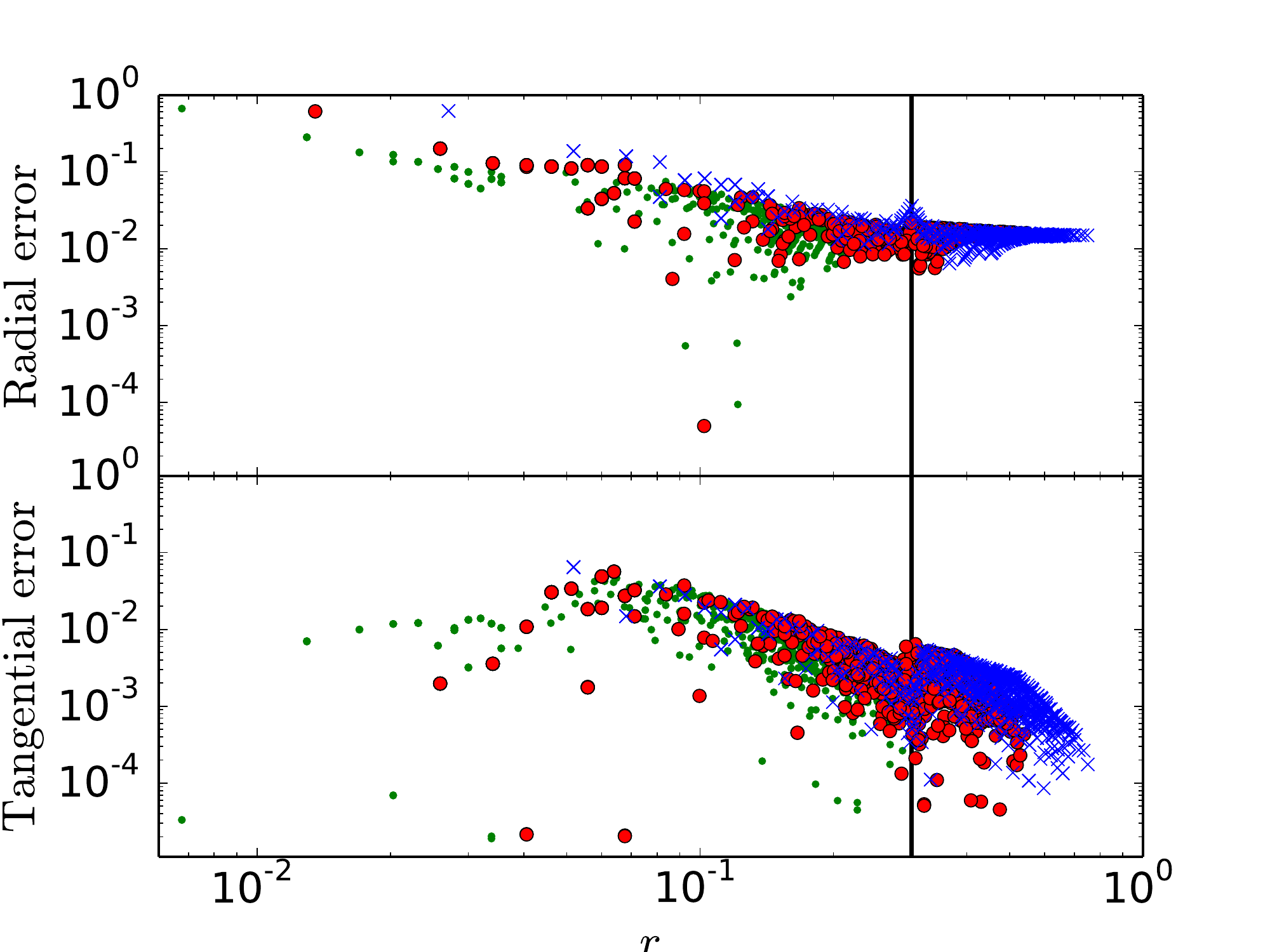}
		\includegraphics[scale=0.44]{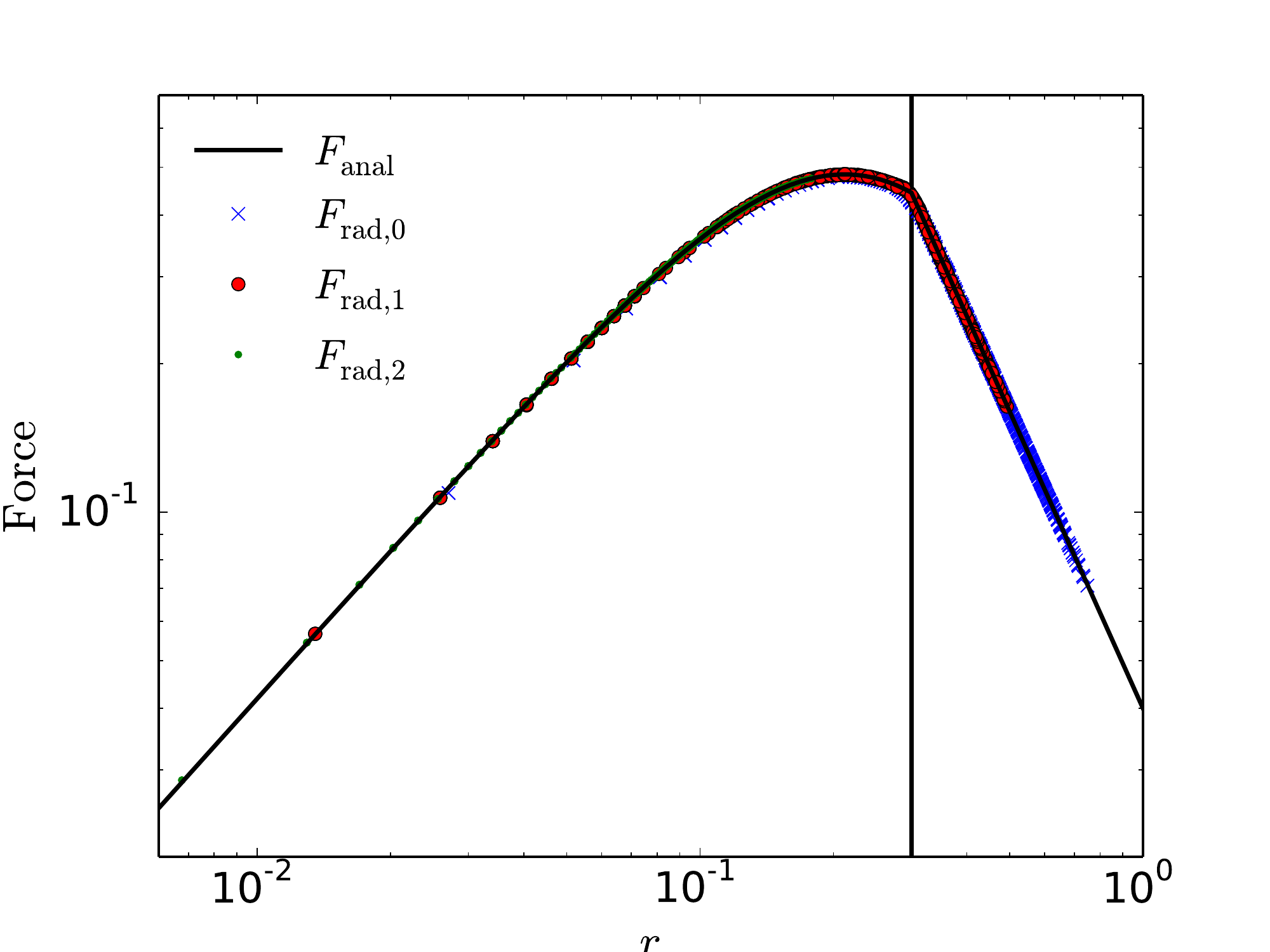}
		\includegraphics[scale=0.44]{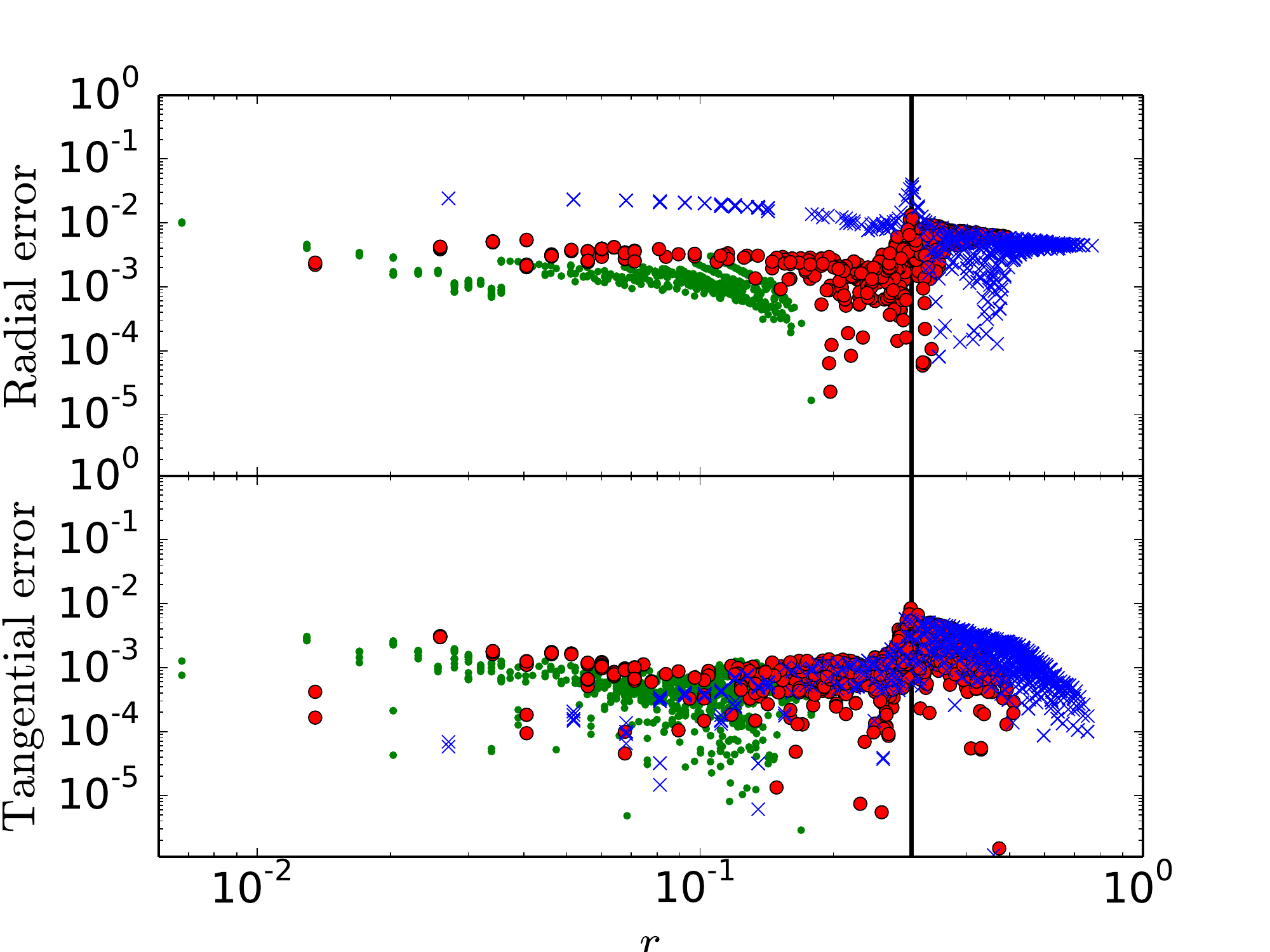}
	\caption{Accelerations (left) and errors (right) computed with the APM solver 
	for the sphere of constant density (top), 
	the isothermal sphere (middle) and the Plummer sphere (bottom). 
	Shown are the analytical force (solid black), the components for the root grid (blue crosses),
	the first level (red filled circles), and the second level of refinement (green dots).
	The radius of the sphere is indicated  by the vertical line.
	Note that the ghost cells
	have been removed from the plots and the analysis.
	\label{fig:spheres-apm}}
	\end{center}
\end{figure*}

The two solvers produce comparable results, therefore we show in Figure~\ref{fig:spheres-apm} 
the results for the APM solver only.
In all three cases the accuracy in both the radial and the tangential directions is at least at the percent level, 
except for the isothermal sphere at small radii.
This is due to the fact that the gravitational acceleration is smoothed near the center 
because it diverges.

\subsection{The sine wave test}
\label{subsec:sinewave}

In order to compare the gravity solvers with periodic boundary conditions, 
we study the case where the gas density is a sinusoidal function:
\begin{equation}
	\rho_{\rm SW}(x,y,z) = 2 + \sin{\left(\frac{2\pi x}{P}\right)},
\end{equation}
which according to Equation~(\ref{eq:PoissonEquation}) leads to the periodic potential:
\begin{equation}
	\Phi_{\rm SW}(x) = -4\pi \left(\frac{P}{2\pi}\right)^2 \sin \left(\frac{2\pi x}{P}\right),
	\label{eq:sinewave}
\end{equation}
where $P$ is the period. 
We use periodic boundary conditions and two static levels of refinement 
with coordinates $[0.34375 ; 0.65625]^3$ (level 1), and $[0.421875 ; 0.578125]^3$ (level 2).
We perform simulations for $P = 0.2$ and $P = 1.0$, and with $64^3$ zones on the root grid.
Solving the Poisson equation with periodic boundary conditions requires that there is no net-mass
in the computational domain. We thus subtract the average density of the system ($\rho_{\mathrm{av}} = 2.0$) from
the \gmf\ before solving Equation~\ref{eq:PoissonFourier}.

\begin{figure*}[!h]
	\begin{center}
		\includegraphics[scale=0.47]{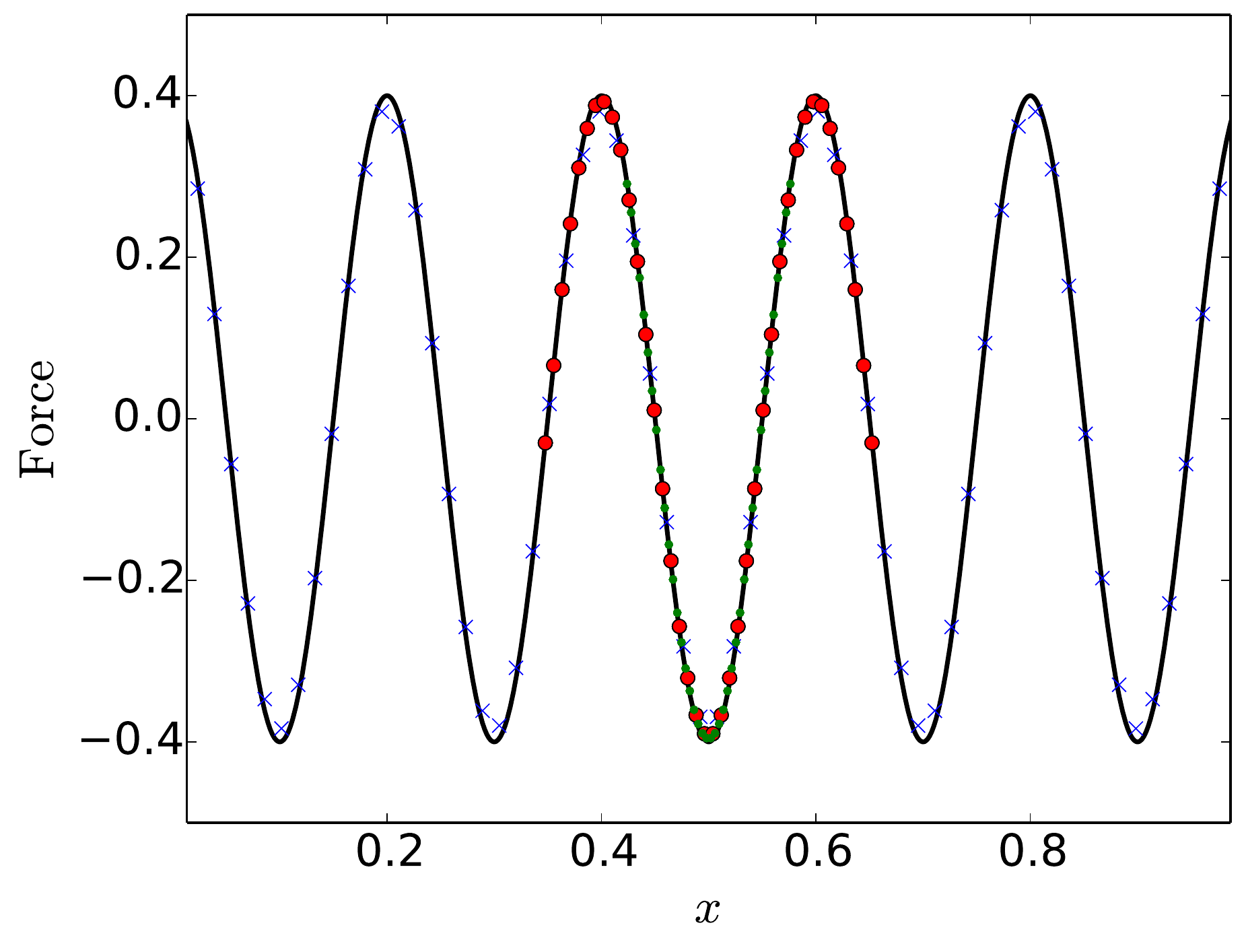}
		\includegraphics[scale=0.47]{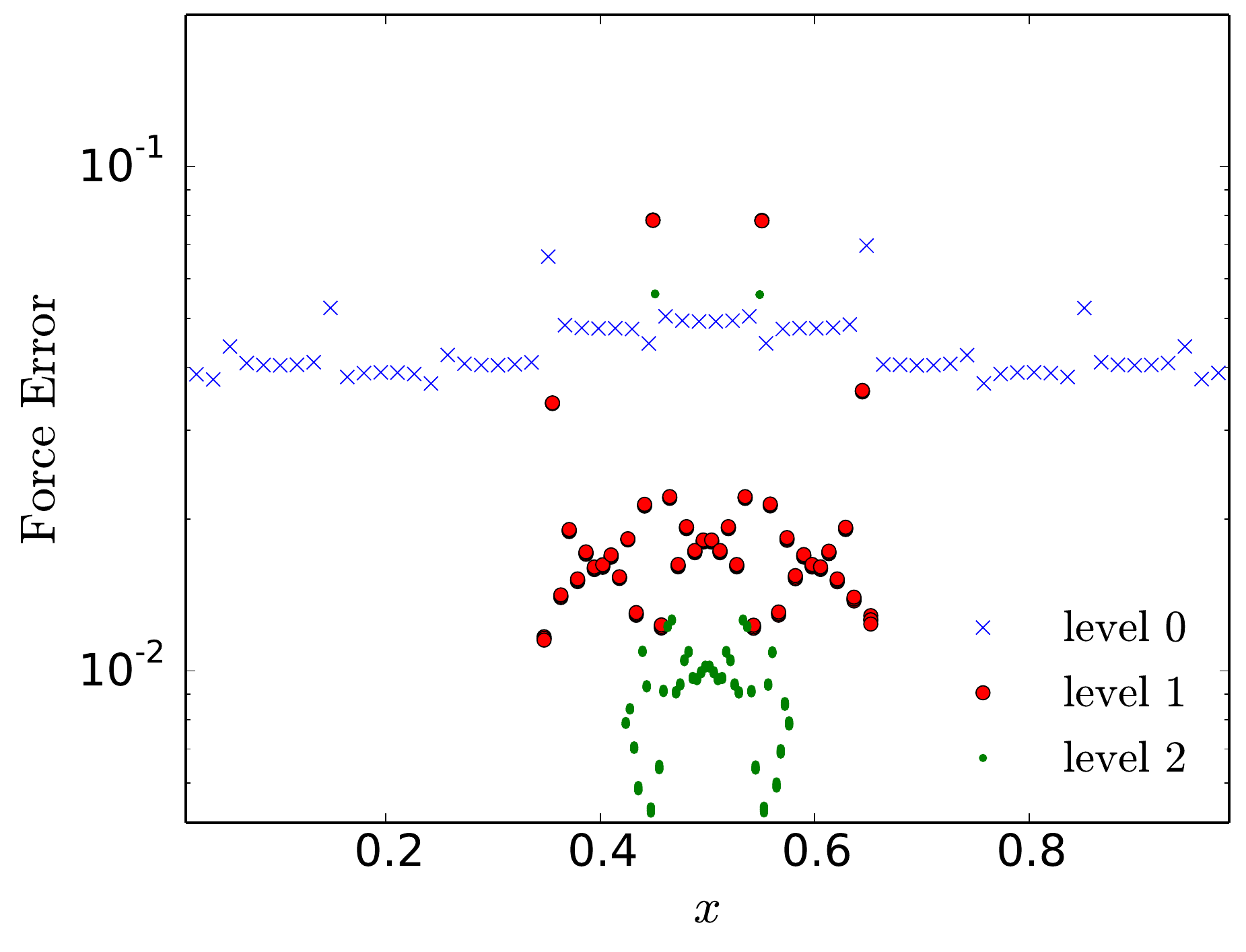}
		\includegraphics[scale=0.47]{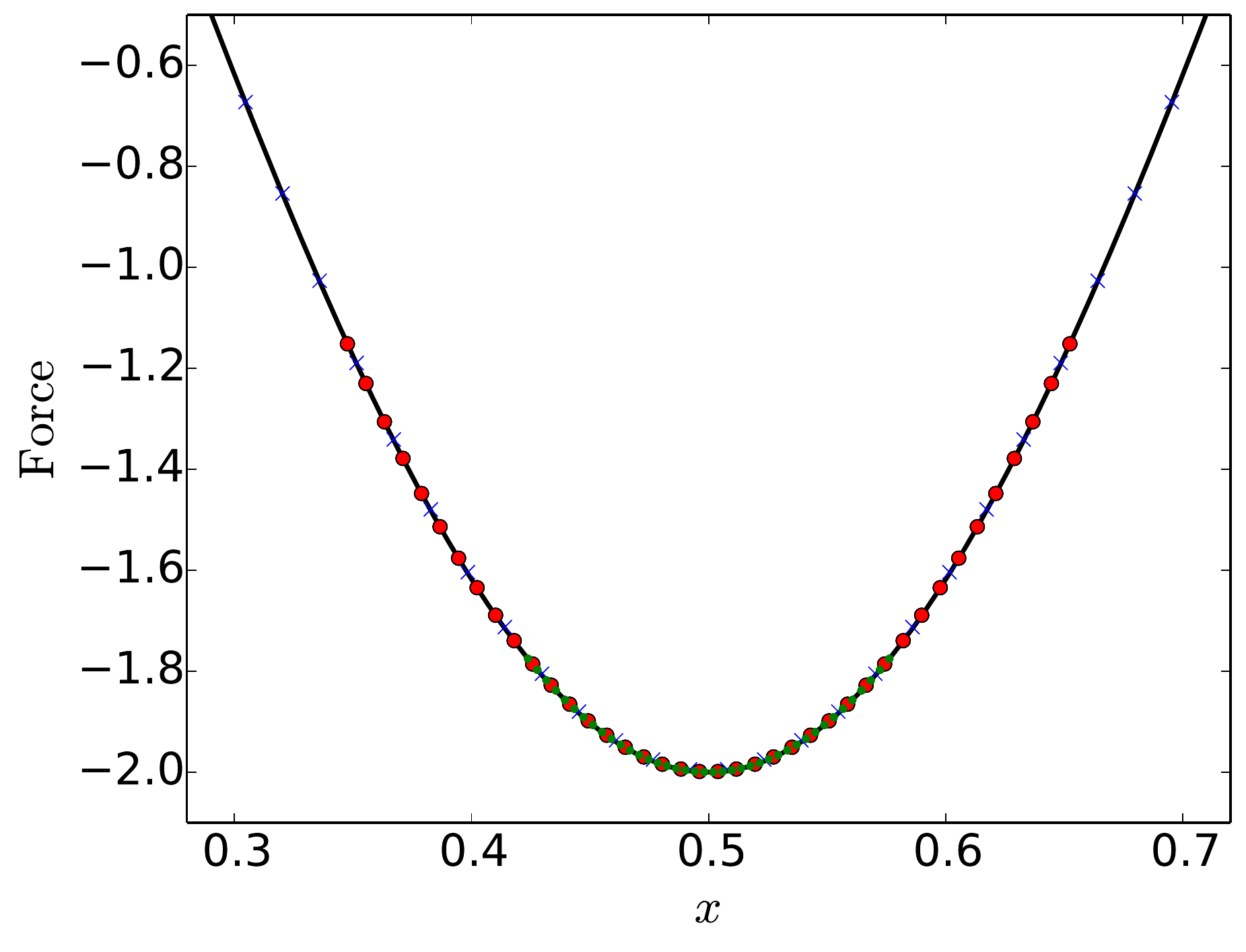}
		\includegraphics[scale=0.47]{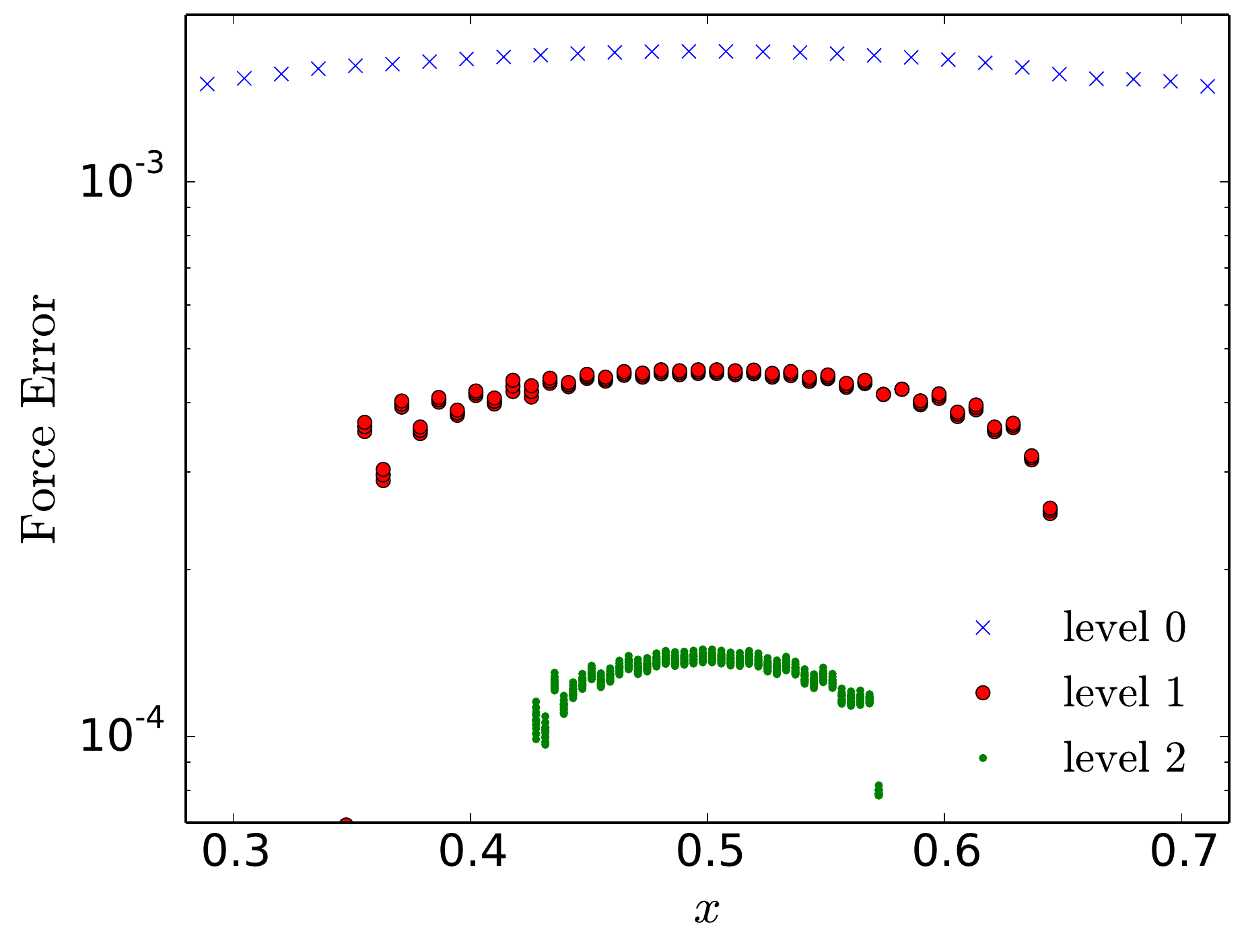}
	\caption{
	Left: comparison between the analytical force (solid black) and the computed force
	using the APM solver for a sine density wave 
	with $P = 0.2$ (rows 1) and  $P = 1.0$ (rows 2).
	The root grid has $64^3$ zones. 
	The colored symbols represent the values of the force calculated along the $x$-axis 
	for the root grid (blue crosses), the first level (red filled circles), 
	and the second level of refinement (green dots) for the active zones.
	We only show the region of interest $x \in [0.3 ; 0.7]$ for the case $P = 1.0$.
	\label{fig:PoissonWave}
	}
	\end{center}
\end{figure*}

The multigrid and the APM solvers give comparable results on this test. 
In Figure~\ref{fig:PoissonWave} we therefore show the computed force and the relative error 
on the force along on the $x$-axis (Equation~\ref{eq:sinewave})for the APM solver only.
For the case $P = 1.0$, the APM solver is quite accurate because the density distribution varies on larger scales.
For the case $P = 0.2$, the accuracy slightly decreases. This behavior is due to the combined facts that
the density varies on smaller scales and that the acceleration is smoothed 
by the TSC deposition and the $S(r)$ function, leading to an effective loss of resolution.

\subsection{The orbit test}
\label{subsec:testorbit}

Finally, we perform a two-body problem with the particles initially in a circular orbit in the $x$-$y$ plane.
The central particle has a mass $M_1 = 1.0$ and is located initially at the center of the domain.
The test particle has a mass $M_2 = 0.1$ and is placed at a distance $a = 0.3$ from the central particle, 
at coordinates $\mathbf{r}_2 = (0.2, 0.5, 0.5)$.
The resolution of the root grid is $16^3$ zones and we study four different cases. 

\subsubsection{Initial configurations}

We investigate four different cases.

\begin{itemize}
	\item Case 1: particles in different levels with subcycling.
	
	In the first case, we allow one level of refinement. Refinement is forced around the central particle such that
the latter lives in level 1. The timestep on the root grid is $dt_0 = 3\times10^{-3}$, and subcycles are allowed such that 
level 1 is evolved with a timestep $dt_1 = dt_0/2$. The system is evolved for 5 orbits with the multigrid solver, 
and for 10 orbits with the APM solver.

	\item Case 2: particles in different levels without subcycling.

	This case is identical to the previous one except that all levels are evolved with a constant timestep $dt = dt_0 = 3\times10^{-3}$.

	\item Case 3: particles in the same level with subcycling.

	Here we allow two levels of refinement and modify the refinement criterion such that both particles
are located in level 2. The binary system is evolved with a constant time step $dt=10^{-3}$ for one orbit with 
the multigrid solver, and 10 orbits with the APM solver. 

	\item Case 4: particles in the same level without subcycling.

	This final case is similar to case 3 but all the levels are updated at a constant timestep $dt = dt_0 = 3\times10^{-3}$.
\end{itemize}

\subsubsection{Results}
\begin{figure*}[!h]
	\begin{center}

		\includegraphics[scale=0.42]{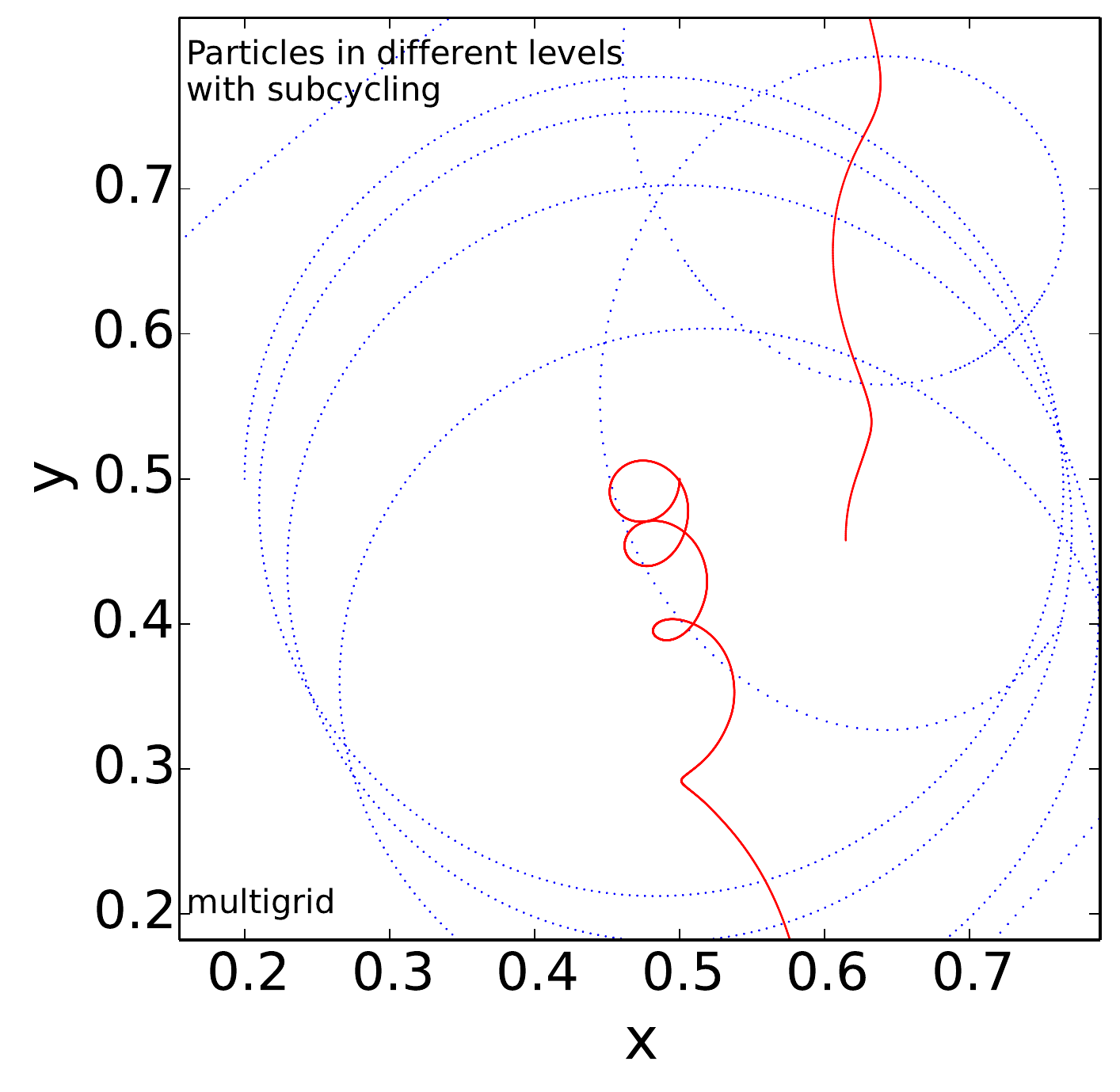}
		\includegraphics[scale=0.42]{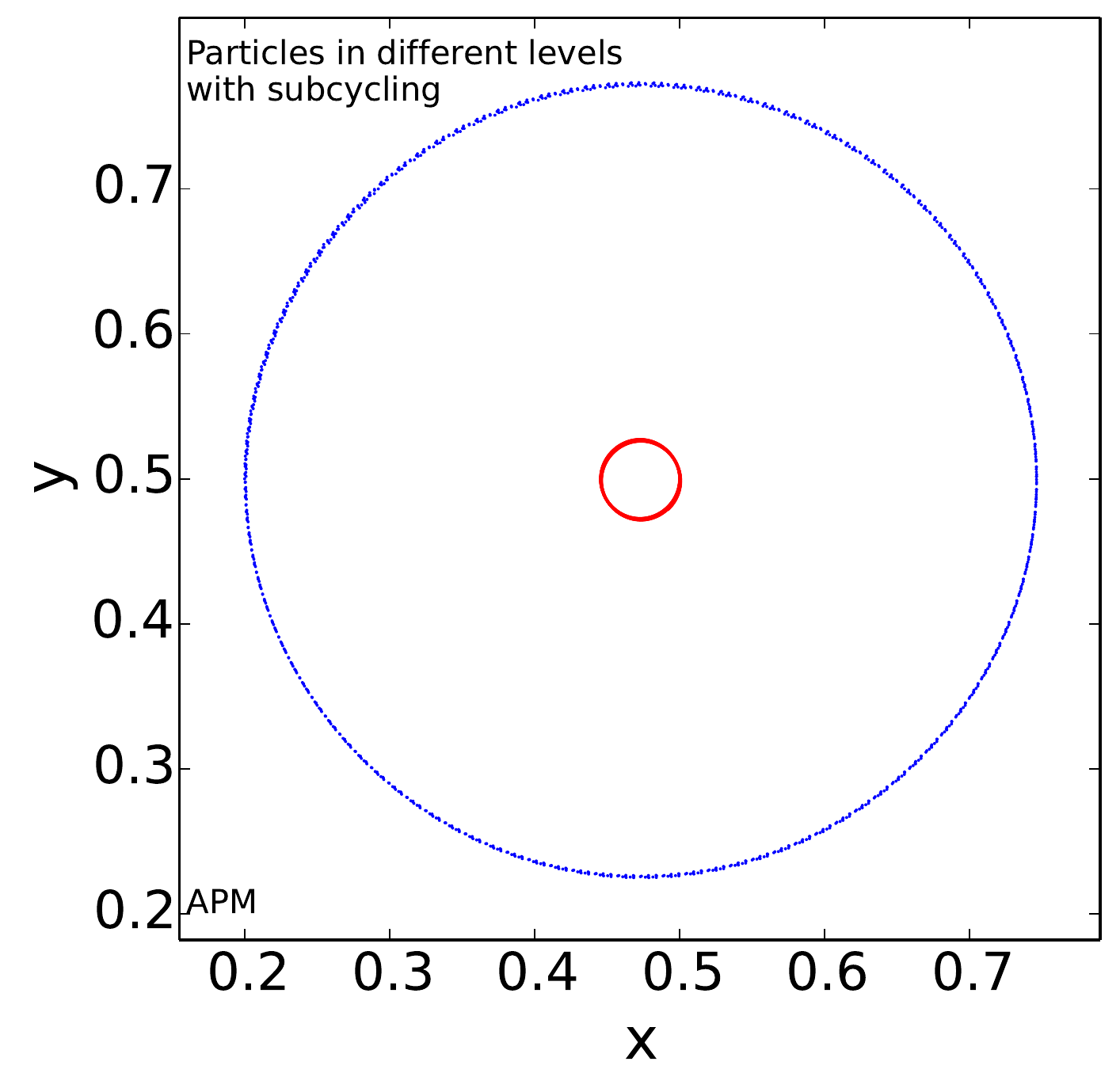}\\
		\includegraphics[scale=0.42]{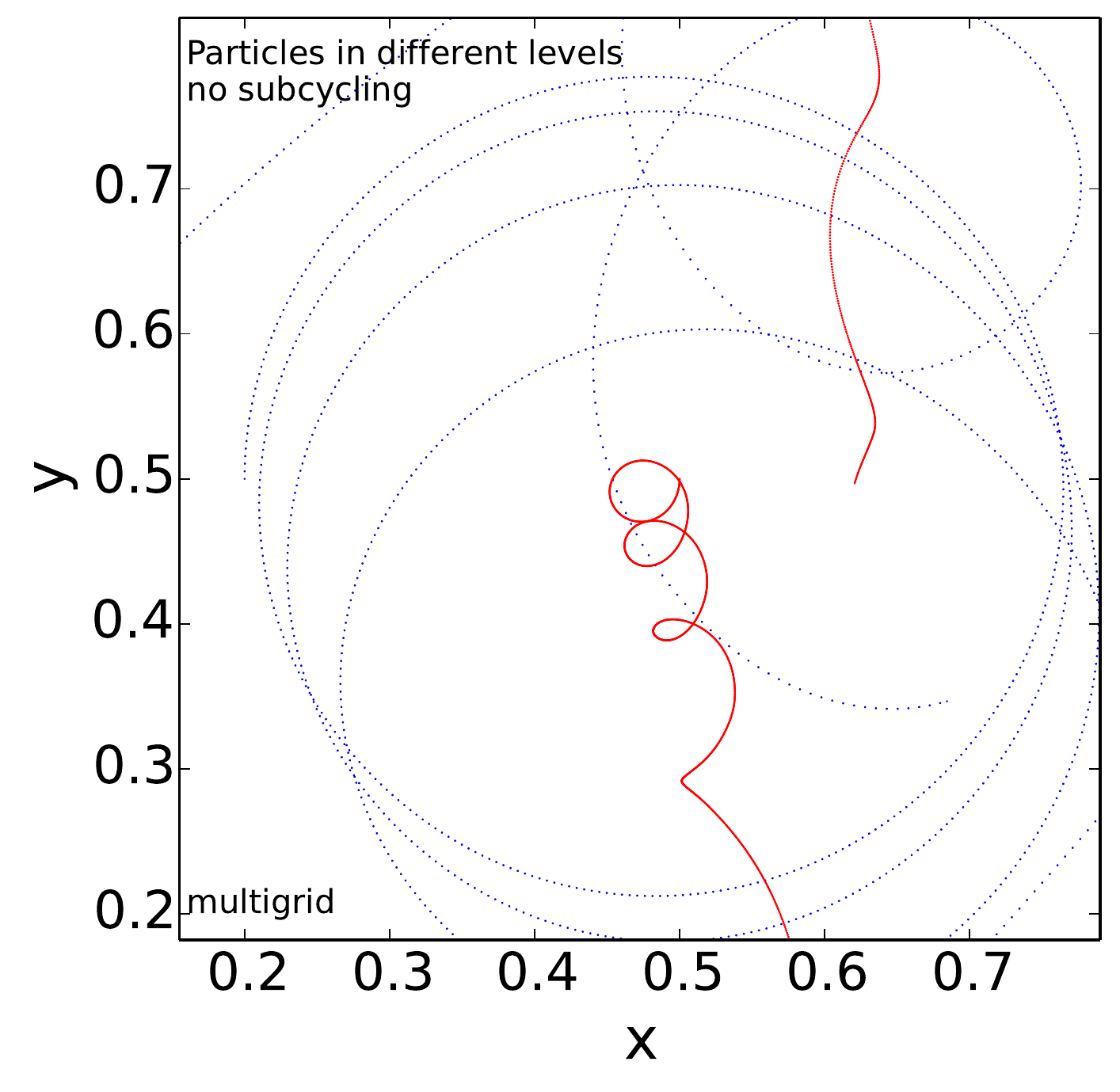}
		\includegraphics[scale=0.42]{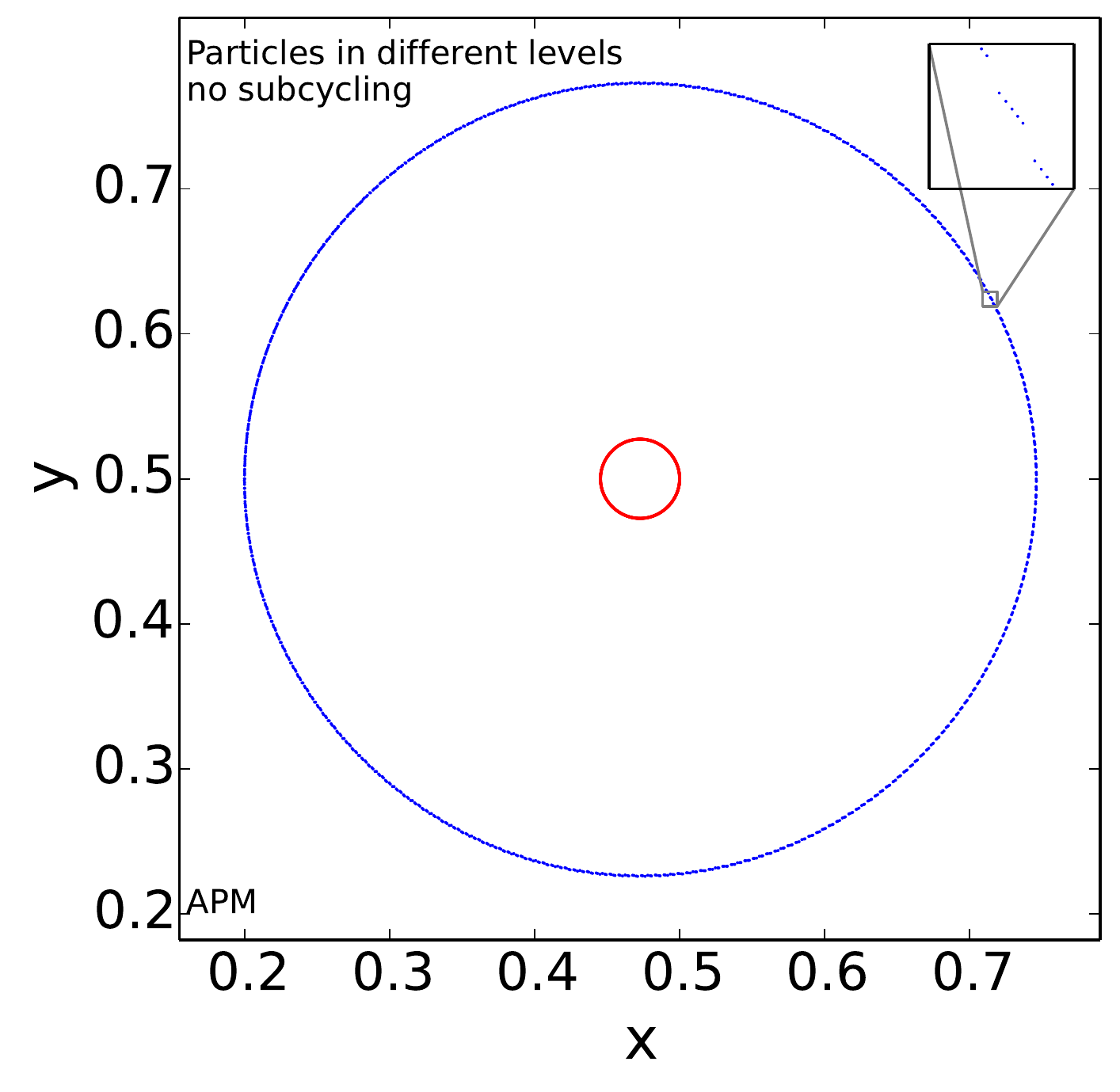}\\
		\includegraphics[scale=0.42]{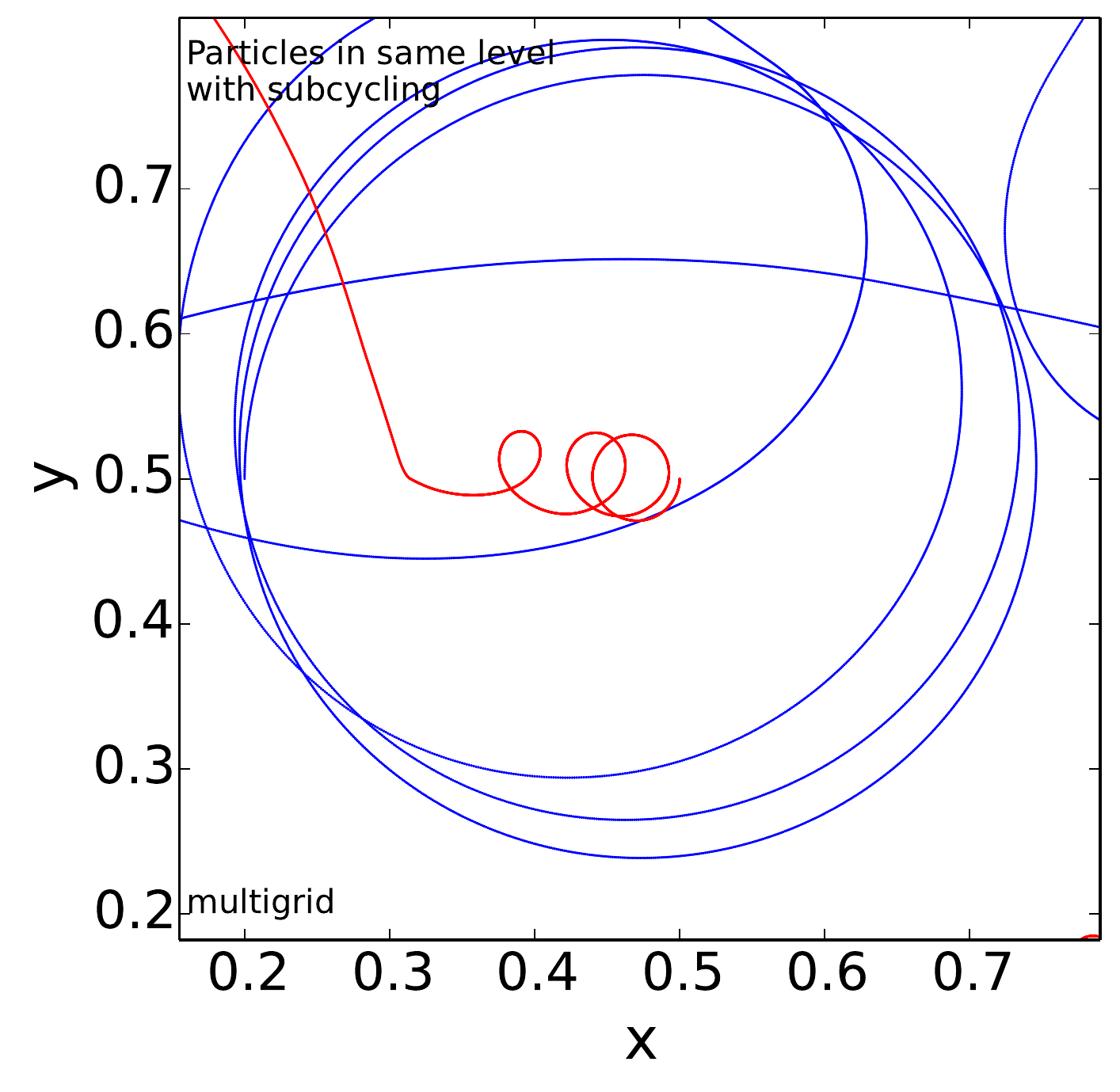}
		\includegraphics[scale=0.42]{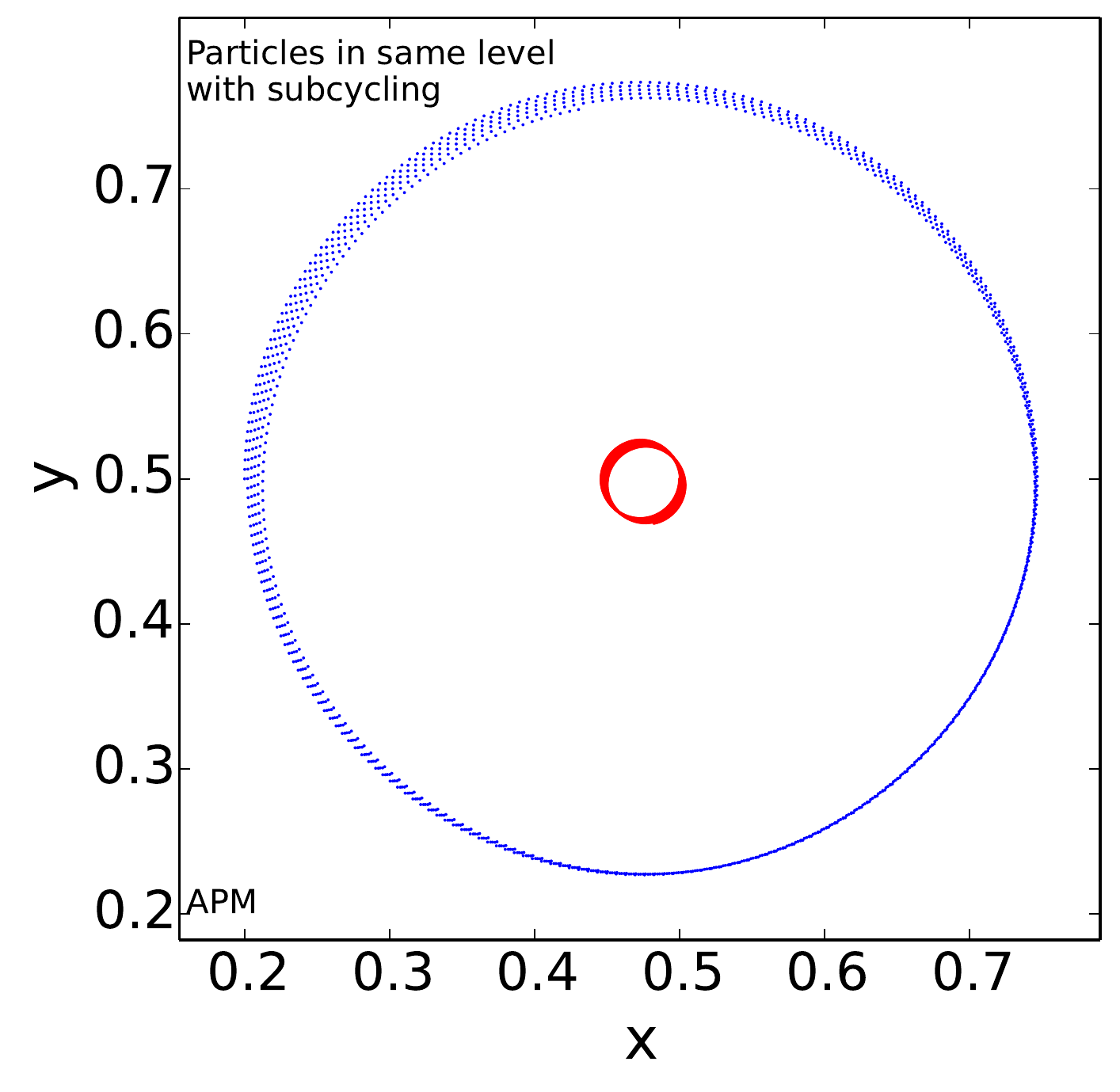}\\
		\includegraphics[scale=0.42]{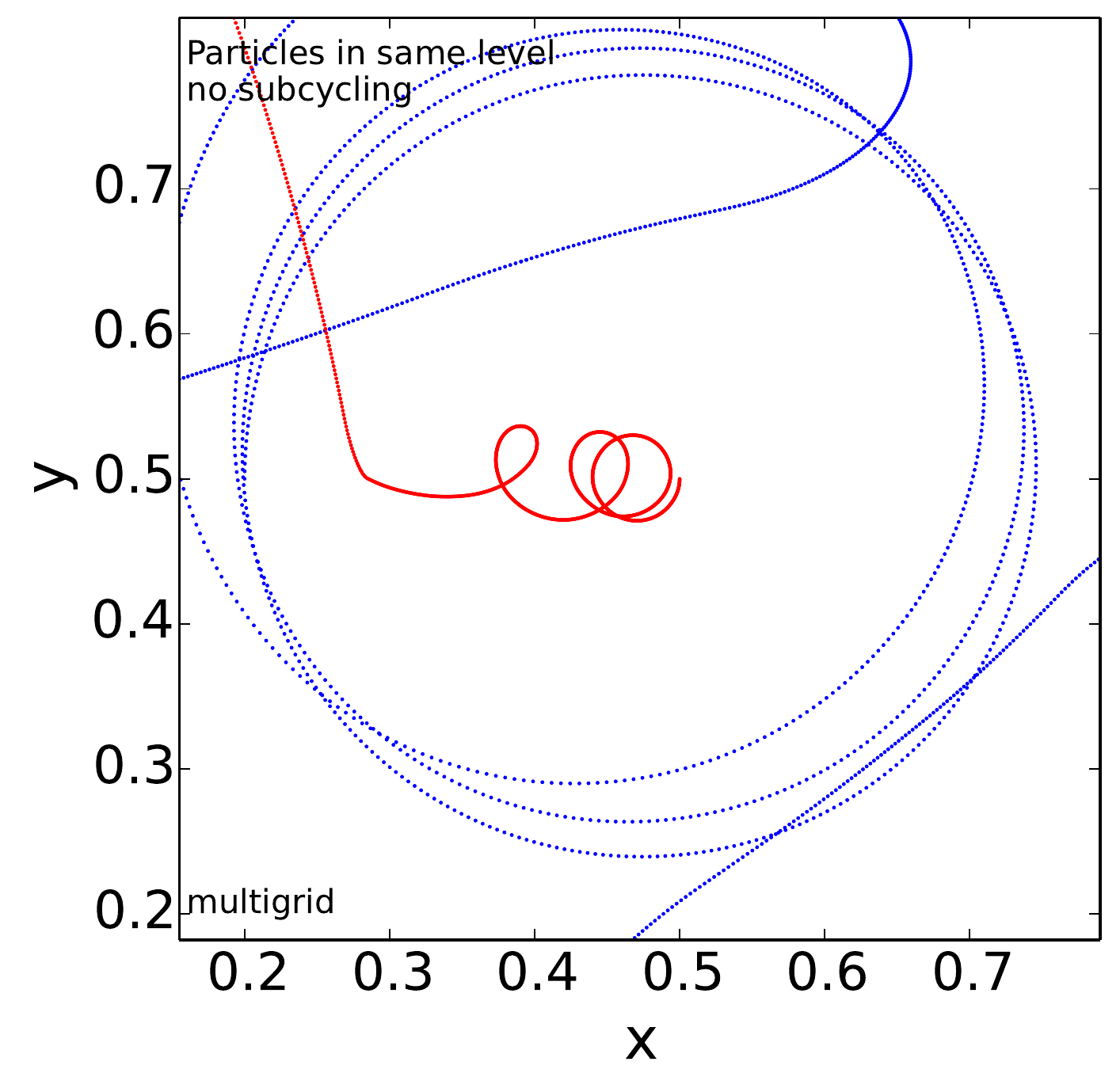}
		\includegraphics[scale=0.42]{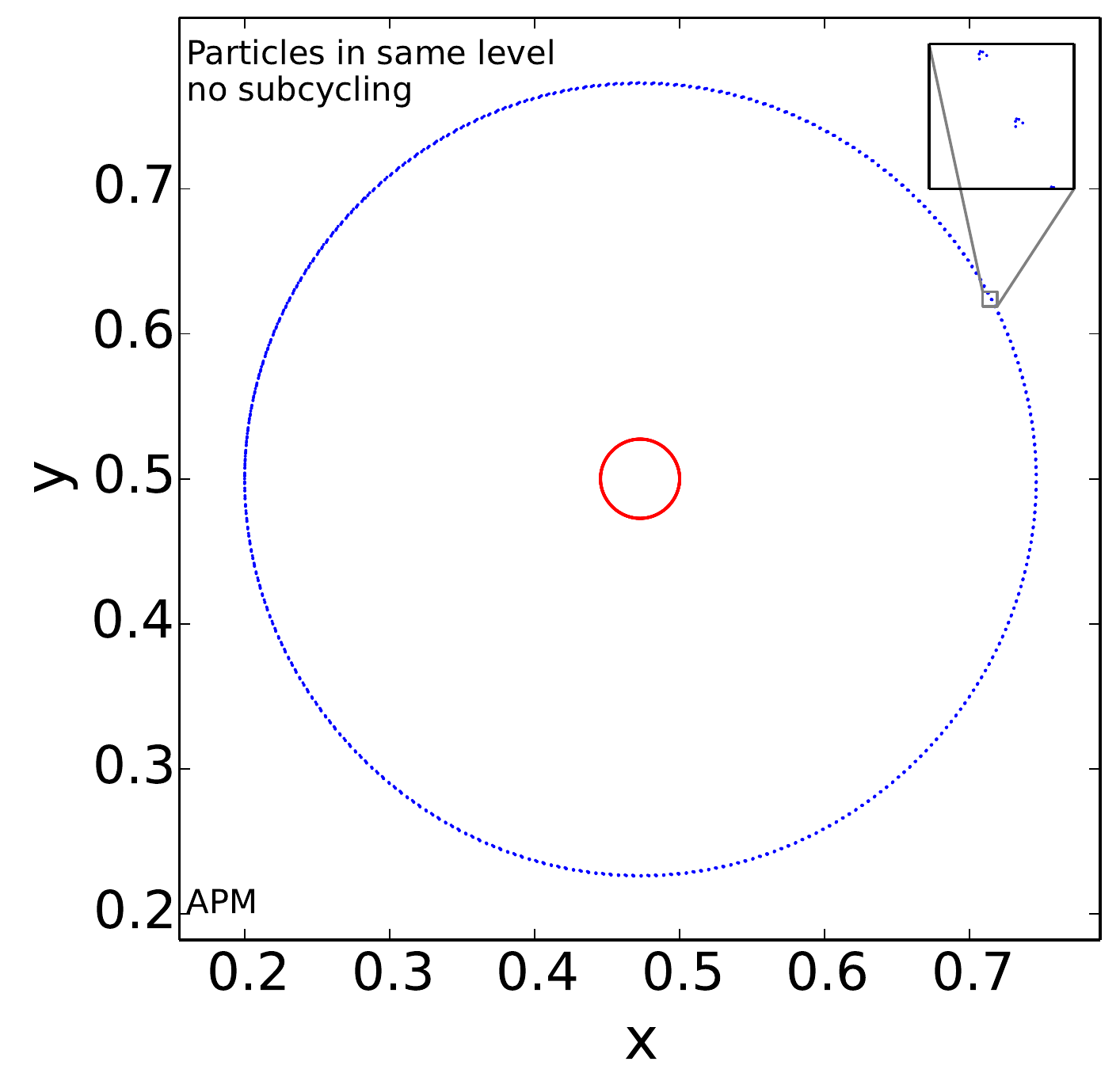}\\
	\caption{Trajectories in the orbital plane of the central particle (red) 
	and the test particle (blue) with the multigrid (left) and the APM solver (right) for the four cases (top to bottom).
	Details about the system configuration are marked on the figures.
	The system has been evolved for five orbits in all cases.
	\label{fig:testorbit}}
	\end{center}
\end{figure*}

The trajectories of the particles are showed for the different cases in Figure~\ref{fig:testorbit}.

In all cases the APM solver gives much more accurate results than the multigrid solver. 
With the multigrid solver, particles have already left the computational domain after less then five orbits 
(Figure~\ref{fig:testorbit}, left column). Note that the particles follow periodic boundary conditions.
One should also highlight that subcycling has very little effect on the multigrid solver.

On the contrary the APM solver yields a much more accurate evolution of the system (Figure~\ref{fig:testorbit}, right column).
Simulations in which subcycling is allowed (cases 1 and 3) show a small resonance of the system: 
the center of mass of the system is shifted during the evolution.
This behavior is most prominent in the cases where both particles are located in the same refined level 
(Figure~\ref{fig:testorbit}, right column, row 3).
Note that while the particles are located in the same level of refinement, 
they are usually not covered by the same grid.
This resonance  is due to the fact that the level where the particles live (level 2) is evolved through
4 subcycles while the root grid level only goes through one, leading to a time-integration inaccuracy.
More quantitatively, the orbital separation of the system has changed by about 3\% at the end of the simulation (case 3).
The center of mass has also moved by about 1\% in the $x$ and $y$-directions.

If subcycling is turned off and all levels are evolved with the same timestep,
the results become much more accurate 
with the APM solver (Figure~\ref{fig:testorbit}, right column, rows 2 and 4). 
Regarding case 4, the orbital separation of the system has  changed by 0.7\% only.
Moreover the center of mass is stable within a relative error $\approx 10^{-5}$.
We thus conclude that the force calculation with the APM solver is quite accurate, 
and that resonance in the orbit can be avoided by
forcing the different levels to be evolved with the same timestepping.

\section{Conclusion}
\label{sec:conclusion}

Modeling accurately self-gravity represents a difficult part of astrophysical simulations.
In this paper we have presented and tested an implementation of the
adaptive particle-mesh solver based on the algorithm 
presented in \cite{Hockney:1989um} and \cite{1991ApJ...368L..23C}. The primary aim of such a 
technique is to provide a more accurate 
computation of the gravitational field at small scales.
The solver has been implemented within the astrophysical code \enzo\ 
but could be used in any other grid-based code with a structured mesh.

We have performed a series of tests in order to examine the accuracy of the APM solver, 
and compared the results with those obtained with the default multigrid solver implemented in \enzo.
For tests in which the gravitating material is distributed over a number of cells (as in, for example,
cosmological simulations) the APM solver and the multigrid solver show similar accuracies.
However, when a small number of particles are used (such that the potentials are very steep),
the APM solver provides much improved results.  In particular, we show that in a two-body problem, 
the code can produce accurate orbits if all the levels are evolved with the same timesteps.
 
An important aspect of the APM algorithm is timestepping. 
If the system is evolved with subcycling, the force components on different scales are evolved at different time,
which leads to some inaccuracy in the time-integration, even though the force calculation on a given level
is computed to high accuracy. This behavior can be seen in particular for the test orbit problem (Section~\ref{subsec:testorbit}).
One solution to get rid of these spurious effects is to disable subcycling and evolve all levels with the same
timestepping.
This will increase the computational cost of the calculation, but it may only be a factor of a few depending on the grid
hierarchy. Such extra accuracy may also only be needed for simulations with only a small number of particles, 
such as the TestOrbit problem. Indeed, the multigrid solver is usually sufficiently accurate with a large number of particles.

Finally, one should recall that the multigrid algorithm relies on interpolating the potential values of the root grid
onto the refined grid, and using a relaxation method until convergence is reached. 
In the APM algorithm, FFTs are used on the refined levels. 
Consequently, the work load on each processor is larger and one expect the APM solver to be slower than the 
multigrid solver. However, the potential values do not need to be communicated between overlapping grids
so communication between processors are decreased. Therefore, we also expect the APM solver
to scale efficiently to a higher number of processors. 
The study of the performance of the APM solver is beyond the scope of this paper, 
but could be investigated in greater detail in the future.
Unlike the multigrid solver, the APM implementation
has not been optimized for parallel performance yet as it does not use the MPI non-blocking communications 
(for more details, see \citealt{2014ApJS..211...19B}).
As an indication, the runtime for the TestOrbit problem with the APM solver is about three to four times as long
as with the multigrid solver.

\section{Acknowledgments}
\label{sec:acks}

J-CP acknowledges funding from NSF grant AST-0607111
and from the Alexander von Humboldt Foundation. 
GLB acknowledges support from NSF grant 1008134 and NASA grant NNX12AH41G.
J-CP thanks Norbert Langer, Mordecai-Mark Mac Low, Falk Herwig, and Orsola De Marco for their support.
We are thankful to the referee for useful comments that improved the clarity of the paper, 
and to Colin P. McNally for suggesting the 
sine wave test presented in this paper.

\bibliographystyle{/Applications/TeX/apj}                       

\end{document}